\documentclass[prd,aps,preprintnumbers,nofootinbib,longbibliography,notitlepage]{revtex4-2}

\usepackage{amsmath,amssymb,mathtools,slashed,bm}
\usepackage{graphicx}
\usepackage{float}
\usepackage{tikz}
\usetikzlibrary{arrows.meta,calc}
\usepackage{booktabs}
\usepackage{xcolor}
\usepackage[colorlinks=true,linkcolor=blue,citecolor=blue,urlcolor=blue]{hyperref}

\graphicspath{{./}}

\newcommand{\Bbar}{\overline B}
\newcommand{\dd}{\mathrm d}
\newcommand{\order}{\mathcal O}
\newcommand{\calE}{\mathcal E}

\newcommand{\calT}{\mathcal T}

\newcommand{\vect}{\bm}
\newcommand{\BTEC}{\mathrm{BTEC}}

\begin{document}

\title{Photon-Tagged Energy Flow in Inclusive Endpoint
$B$ Decays}

\author{Shuai Zhao}
\email{zhaos@tju.edu.cn}
\affiliation{Department of Physics and Center for Joint Quantum Studies,
School of Science, Tianjin University, Tianjin 300350, China}

\date\today

\begin{abstract}
We introduce a $B$-decay tagged energy correlator (BTEC) to resolve the
angular structure of energy flow in inclusive $B$ decays in the endpoint
region, focusing on the direct-photon contribution to
$\Bbar\to X_s\gamma$.  At leading power and at the natural collinear angular
scale $\tau\sim s\sim Q\Lambda_{\rm QCD}$, we derive a factorization relation
involving the standard hard coefficient and $B$-meson shape function together
with a new measured quark jet function; no new leading-power nonperturbative
function is introduced.  We calculate the measured jet function at one-loop
accuracy and verify that its angular integral reproduces the standard
inclusive quark jet function.  For the central
Bosch--Lange--Neubert--Paz parameter set, an illustrative benchmark gives
$3 - 16\%$ migration outside the fixed angular cuts
$\tau_c=2 - 4~{\rm GeV}^2$.  With an independently constrained shape
function, the angular cumulatives provide a closure test of leading-power
endpoint factorization and are sensitive to direct-$O_7$ power corrections
and resolved-photon effects.  The BTEC thereby adds information on the
angular structure of the inclusive recoil jet beyond the ordinary photon
spectrum and may improve signal--background discrimination when their
energy-flow profiles differ.
\end{abstract}

\maketitle

\section{Introduction}

Inclusive heavy-flavor decays provide precision tests of the heavy-quark expansion and determine fundamental flavor parameters.
For sufficiently inclusive observables, a local operator product expansion organizes the decay rate in powers of $1/m_b$~\cite{Manohar:2000dt}.
Near a kinematic endpoint, however, the final-state hadronic system has a large energy and a parametrically smaller invariant mass.
The local expansion is then replaced by a nonlocal factorization theorem involving hard, collinear, and soft degrees of freedom
\cite{Neubert:1993um,Bauer:2000ew,Bauer:2001yt,Bosch:2004th,Lange:2005yw,Paz:2022opz}.

The radiative decay $\Bbar\to X_s\gamma$ is particularly well suited to differential studies of the endpoint jet.
The energetic photon is experimentally clean, tags the recoil direction event by event, and determines the exact small light-cone component $P_+=m_B-2E_\gamma$.
Its energy satisfies $2E_\gamma=Q-P_+$ and approaches the fixed large component $Q=P_-=m_B$ only in the endpoint limit.
The ordinary photon spectrum probes the light-cone momentum distribution of the heavy quark through the leading $B$-meson shape function.
Its perturbative description and the residual theory uncertainties in inclusive
radiative decay continue to be refined~\cite{Misiak:2020vlo,Gunawardana:2019gep}.
However, it does not by itself resolve how the recoil energy is distributed in angle inside the inclusive strange final state. This missing angular information motivates us to introduce the $B$-decay tagged energy correlator (BTEC).

An event shape observable, namely the energy-energy correlator (EEC)~\cite{Basham:1979gh,Basham:1977iq,Basham:1979gh,Basham:1978bw}(for a review, see Ref.~\cite{Moult:2025nhu} and references therein), was also proposed for heavy flavor jets~\cite{Craft:2022kdo,Chen:2024nfl,Barata:2025uxp,ALICE:2025igw} as well as $B$ decays long ago~\cite{Luke:1993pg}.
The one-point energy-flow observable itself dates to
Ref.~\cite{Sterman:1975xv}, while modern formulations connect energy
correlators to light-ray operators and their operator product expansion
~\cite{Hofman:2008ar,Kologlu:2019mfz,Chang:2020qpj,Chen:2021gdk,Chen:2023zzh}.
Pairwise EECs correlate two energy-flow insertions in the final state and provide infrared-safe probes of QCD radiation
\cite{Basham:1978bw,Basham:1978zq,Belitsky:2001ij,Dixon:2019uzg,Chen:2020vvp}.
Recent progresses include analytic higher-point correlators, logarithmic
resummation, and studies of the perturbative--nonperturbative transition
~\cite{Chen:2022jhb,Yang:2022tgm,Chen:2023zlx,Schindler:2023cww,
Liu:2024lxy,Chen:2024nyc,Lee:2024esz,Kang:2024dja,Herrmann:2025fqy}.

In $\Bbar\to X_s\gamma$, however, the observed  photon fixes one reference direction, while the radiative two-body partition supplies the fixed hard scale $Q=P_-=m_B$. 
The use of a tagged energy correlator, rather than a conventional EEC, follows from this event geometry.
Only one hadronic energy-flow insertion is therefore needed to resolve the recoil jet relative to the tag~\cite{Ricci:2022htc}.
Accordingly, the BTEC is a photon-tagged one-point energy correlator rather
than a pairwise hadronic EEC.  It measures the energy density of the entire
inclusive state $X_s$ relative to the exact photon recoil axis.  Unlike jet
shapes and angularities, it requires neither jet reconstruction nor an externally imposed jet-radius parameter, while preserving the full angular dependence instead of reducing it to a single weighted moment~\cite{Kang:2017mda,Cal:2019hjc,Lee:2009cw,Larkoski:2014uqa}.
Its closest jet-based analogue is the one-point correlator inside a
reconstructed jet, which instead targets final-state TMD fragmentation
\cite{Mi:2025abd,Makris:2018npl}.
Related one-point energy-flow observables have been developed for
lepton--hadron scattering~\cite{Cao:2023oef,Fu:2025hpc,Kang:2026hig}, while
azimuthal, transverse, and vector-boson-tagged correlators connect energy
flow to TMD dynamics~\cite{Kang:2023big,Gao:2023ivm,Kang:2024otf,Cao:2025icu}.

This photon-tagged one-point geometry leads directly to the factorization
problem we are going to address in this work.  We focus on the direct $O_7$ contribution
to $\Bbar\to X_s\gamma$ at the natural collinear angular scale and derive the
leading-power factorization relation,
by showing that the measurement acts only
on the collinear sector at leading power and then using the
Bauer--Pirjol--Stewart (BPS) field redefinition~\cite{Bauer:2001yt} to separate the
collinear and soft matrix elements.  
We will demonstrate that the tagged angular
measurement preserves the standard leading $B$-meson shape function, while
the new angular dependence is encoded in a measured quark jet function.  We
define this function as a cut operator matrix element, calculate it at one
loop as a distribution in $(s,\tau)$, where $\tau$ is a dimension-two angular variable, and verify that its inclusive
projection reproduces the standard quark jet function. 
The factorization relation is restricted to the direct $O_7$ contribution in the
endpoint region $P_+\sim\Lambda_{\rm QCD}$ and at the natural collinear
angular scale $s\sim\tau\sim Q\Lambda_{\rm QCD}\ll Q^2$, with the measurement
made relative to the fixed photon-defined axis.  Resolved-photon effects
enter at subleading power and require additional soft and jet functions
\cite{Benzke:2010js,Benzke:2010tq,Benzke:2022vki,Bartocci:2024bbf}, while wide-angle energetic radiation
requires matching beyond the leading collinear description.  At
$\tau\sim\Lambda_{\rm QCD}^2$, soft--collinear transverse-momentum balance
becomes leading and a TMD-like soft function is required. This
parametrically smaller-angle regime lies beyond the present analysis.

The BTEC  provides an angularly
resolved view of endpoint dynamics that the ordinary inclusive spectrum
projects onto a one-dimensional convolution.  The leading
shape function remains one-dimensional and universal, but the measured jet
kernel maps the same nonperturbative input into a family of observable
$(E_\gamma,\tau)$ distributions.  This allows the data to
test the leading-power description and constrain its corrections rather than
simply refit $\widehat S_B$.  
Once $\widehat S_B$ is calibrated with the photon spectrum or lattice
simulations, BTEC measurements can test power corrections to radiative
endpoint factorization.  The same leading shape function also enters
semileptonic endpoint decay, and the measured-jet construction admits a
corresponding extension to $\Bbar\to X_u\ell\bar\nu$.  The radiative channel
can therefore provide a calibrated baseline for testing the shape-function
universality before the angular information is used in an inclusive
$|V_{ub}|$ analysis~\cite{Hashimoto:2017wqo,Barone:2023tbl,Belle:2021eni}.

The rest of the paper is organized as follows.
Sec.~\ref{sec:def} defines the observable and its kinematics.
Sec.~\ref{sec:fact} gives the leading-power factorization derivation as a self-contained part of the main text.
Secs.~\ref{sec:jet} and \ref{sec:rg} present the one-loop result and the renormalization-group evolution of the measured jet function, respectively.
Sec.~\ref{sec:cumulative} constructs fixed-cut cumulatives, presents an
illustrative numerical benchmark, and develops the closure test, experimental
feasibility, and semileptonic extension. A summary of this work is presented in Sec.~\ref{sec:summary}.

\section{Endpoint kinematics and observable}
\label{sec:def}

We work in the $B$-meson rest frame, $v^\mu=(1,\vect 0)$ with $v$ denoting the velocity of the $B$-meson.
For a hadronic final state with total momentum $P_X^\mu$, define $P_+ = E_X-|\vect P_X|$ and $P_- = E_X+|\vect P_X|$.
We use $Q\equiv P_-$ for the large light-cone component.
The shape-function region is characterized by $P_+\sim\Lambda_{\rm QCD}$, $Q\sim m_b$, and $M_X^2=QP_+\sim m_b\Lambda_{\rm QCD}$ with $M_X$ denoting the invariant mass of the hadronic final state.

For the radiative decay, the observed photon direction defines the recoil axis $\hat n_J=-\hat n_\gamma$.
Four-momentum conservation in the $B$-meson rest frame and $q^2=0$ give
\begin{align}
  E_X&=m_B-E_\gamma,~~|\vect P_X|=E_\gamma,
~~  Q=P_-=m_B, ~~P_+=m_B-2E_\gamma.
  \label{eq:recoil}
\end{align}
Note that near the endpoint, one has $ 2E_\gamma=Q-P_+\simeq Q$.
Throughout the analysis, $Q$ denotes the exact large component $P_-=m_B$ instead of the event-dependent quantity $2E_\gamma$.
The partonic hard matching scale remains of order $m_b$, and its distinction from the hadronic quantity $Q=m_B$ belongs to the power expansion.

\begin{figure}[t]
  \centering
  \includegraphics[width=0.7\linewidth]{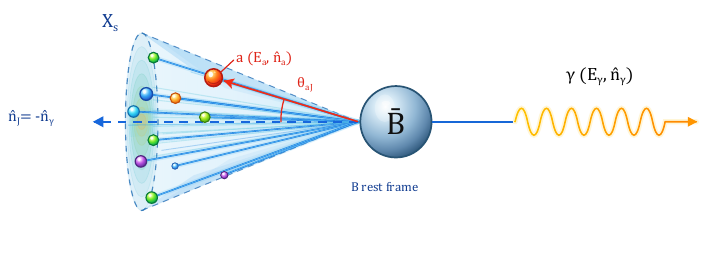}
  \caption{Photon-tagged energy flow in $\Bbar\to X_s\gamma$.
  The observed photon fixes the recoil axis $\hat n_J=-\hat n_\gamma$.
  Each hadron contributes with energy weight $E_a/E_X$ at the angular variable
  $\tau_a=Q^2(1-\hat n_J\cdot\hat n_a)/2$.
  The exact radiative kinematics gives $P_-=Q=m_B$, while the endpoint
  region has $P_+\sim\Lambda_{\rm QCD}$ and $Q\sim m_b$.}
  \label{fig:btec}
\end{figure}

The basic geometry is shown in Fig.~\ref{fig:btec}.
For each hadron $a\in X$, we define the dimension-two angular variable
\begin{align}
  \tau_a
  =Q^2\frac{1-\hat n_J\cdot\hat n_a}{2}
  =Q^2\frac{1+\hat n_\gamma\cdot\hat n_a}{2}.
  \label{eq:tau}
\end{align}
Its physical support is $0\le\tau_a\le Q^2$.
For a narrow particle at angle $\theta_{aJ}$ from the recoil axis,
$\tau_a\simeq Q^2\theta_{aJ}^2/4$.
Here $\hat n_a$ is the physical direction of particle $a$, while $\hat n_J$
is fixed exactly by the tagged photon.

For each hadronic final state $X$, define the normalized event-level
energy-flow measurement
\begin{align}
  \calT_X(\tau)
  =\frac{1}{E_X}\sum_{a\in X}E_a\,\delta(\tau-\tau_a),
  \qquad
  \int_0^{Q^2}\dd\tau\,\calT_X(\tau)=1,
  \label{eq:eventEnergyFlow}
\end{align}
for a fully visible hadronic final state.
In the endpoint region,
\begin{align}
  E_X=\frac{P_-+P_+}{2}
  =\frac{Q}{2}\left[1+\order\!\left(\frac{\Lambda_{\rm QCD}}{Q}\right)\right].
  \label{eq:EXleading}
\end{align}
Consequently, the leading collinear measurement weights are the large light-cone energy fractions of the final-state particles.
The same event-level measurement can also be expressed with the energy-flow operator through
\begin{align}
	\calT_X(\tau)
	=\frac{1}{E_X}\int\dd\Omega\,
	\delta\!\left(\tau-Q^2\frac{1-\hat n_J\cdot\hat n}{2}\right)
	\langle X|\calE(\hat n)|X\rangle,
	\label{eq:flowform}
\end{align}
where the energy-flow operator acts on an asymptotic state as
\begin{align}
  \calE(\hat n)|X\rangle
  =\sum_{a\in X}E_a\,
  \delta^{(2)}(\hat n-\hat n_a)|X\rangle.
\end{align}
The energy weight makes the measurement soft safe, while the sum of daughter energies makes it inclusive under a collinear splitting.

Before continuing, we make a remark on the moment of $\calT_X$. The first angular moment is fixed almost entirely by event kinematics.
In the massless-particle convention used in the leading-power partonic calculation,
$\sum_a E_a\hat n_a=\vect P_X$ and $\hat n_J=\vect P_X/|\vect P_X|$.
Therefore,
\begin{align}
  \int_0^{Q^2}\dd\tau\,\tau\,\calT_X(\tau)
  &=\frac{Q^2}{2E_X}
  \left(E_X-\hat n_J\cdot\sum_a E_a\hat n_a\right)
  =\frac{Q^2P_+}{Q+P_+}
  =\frac{M_X^2}{1+P_+/Q}
  =M_X^2+\order(\Lambda_{\rm QCD}^2),
  \label{eq:firstMomentIdentity}
\end{align}
where $Q=P_-$ and $M_X^2=QP_+$ are exact.
Only the last equality expands the exact event-level result in $P_+/Q$.
This identity indicates that a full-angle positive moment is not a clean probe of the collinear angular profile.
A wide-angle soft particle has $E_s/E_X\sim\Lambda_{\rm QCD}/Q$ but
$\tau_s\sim Q^2$, and hence contributes at order $Q\Lambda_{\rm QCD}$, the same order as $M_X^2$.
The first moment is thus essentially a rewriting of the hadronic invariant mass and supplies a nontrivial consistency check on a fixed-order calculation.
Genuinely new angular information resides instead in the differential distribution or in a bounded cumulative for which the collinear expansion is uniform.

The observable named as the BTEC in this work is the photon-energy and
angular double-differential decay spectrum.  At fixed photon energy, let
$q^\mu=E_\gamma(1,\hat n_\gamma)$ and insert
Eq.~\eqref{eq:eventEnergyFlow} into the state-resolved rate, one has
\begin{align}
  \frac{\dd^2\Gamma_{\BTEC}^{s\gamma}}
       {\dd E_\gamma\,\dd\tau}
  &\equiv
  \frac{E_\gamma}{2(2\pi)^3}
  \frac{1}{2m_B}
  \sum_{X,\lambda_\gamma}
  \int\dd\Omega_\gamma\,\dd\Phi_X\,
  (2\pi)^4\delta^{(4)}(p_B-q-P_X)
  |\mathcal A(\Bbar\to X_s\gamma)|^2
  \calT_X(\tau),
  \label{eq:rateDefinition}
\end{align}
where $E_\gamma/[2(2\pi)^3]$ is the massless one-photon phase space per
unit energy.  The photon direction is integrated over the full solid angle;
detector acceptance and fiducial photon cuts are not part of the theoretical
definition and can be applied separately.
Thus $\calT_X(\tau)$ is the measurement function assigned to one hadronic
state, whereas $\dd^2\Gamma_{\BTEC}^{s\gamma}/(\dd E_\gamma\dd\tau)$ is the
ensemble observable to be measured in finite $(E_\gamma,\tau)$ bins.

\section{Leading-power factorization}
\label{sec:fact}

We now derive the leading-power factorization relation for the direct-photon contribution with the strategy of effective field theory.
The derivation starts from the full-QCD measured decay tensor and reduces its current, state sum, kinematic constraint, and energy-flow insertion one by one.

\subsection{Measured decay tensor in full QCD}

The ordinary photon spectrum and the BTEC spectrum differ only by the energy-flow insertion.
They are defined by
\begin{align}
  \frac{\dd\Gamma^{s\gamma}}{\dd E_\gamma}
  &=\frac{E_\gamma}{2(2\pi)^3}\frac1{2m_B}
  \sum_{X,\lambda_\gamma}
  \int\dd\Omega_\gamma\,\dd\Phi_X\,
  (2\pi)^4\delta^{(4)}(p_B-q-P_X)
  |\mathcal A(\Bbar\to X_s\gamma)|^2,
  \nonumber\\
  \frac{\dd^2\Gamma_{\BTEC}^{s\gamma}}
       {\dd E_\gamma\dd\tau}
  &=\frac{E_\gamma}{2(2\pi)^3}\frac1{2m_B}
  \sum_{X,\lambda_\gamma}
  \int\dd\Omega_\gamma\,\dd\Phi_X\,
  (2\pi)^4\delta^{(4)}(p_B-q-P_X)
  |\mathcal A(\Bbar\to X_s\gamma)|^2
  \calT_X(\tau).
  \label{eq:fullQCDRatesPair}
\end{align}
Integrate over the angular variable leads to
\begin{align}
  \int_0^{Q^2}\dd\tau\,
  \frac{\dd^2\Gamma_{\BTEC}^{s\gamma}}
       {\dd E_\gamma\dd\tau}
  =\frac{\dd\Gamma^{s\gamma}}{\dd E_\gamma}.
  \label{eq:fullQCDInclusiveRelation}
\end{align}
The task of factorization is therefore to determine how the additional operator
$\calT_X(\tau)$ resolves the endpoint final state without disturbing the known inclusive structure.

Before performing an endpoint expansion, contract the photon field in the weak Hamiltonian with the observed photon state and define the full-QCD transition current by
\begin{align}
  \langle X_s\gamma(q,\lambda)|\mathcal H_{\rm eff}|\Bbar\rangle
  =\epsilon_\mu^*(q,\lambda)
  \langle X_s|\mathcal J_\gamma^\mu(q)|\Bbar\rangle.
  \label{eq:fullQCDCurrentDefinition}
\end{align}
Eq.~\eqref{eq:rateDefinition} can then be reorganized as
\begin{align}
  \frac{\dd^2\Gamma_{\BTEC}^{s\gamma}}
       {\dd E_\gamma\dd\tau}
  &=\frac{E_\gamma}{2(2\pi)^3}
  \int\dd\Omega_\gamma\,
  \sum_\lambda
  \epsilon_\mu^*(q,\lambda)\epsilon_\nu(q,\lambda)
  W_{\BTEC}^{\mu\nu}(\tau;q),
  \label{eq:fullQCDRateTensor}
\end{align}
where the measured hadronic tensor is
\begin{align}
  W_{\BTEC}^{\mu\nu}(\tau;q)
  &=\frac1{2m_B}\sum_X\int\dd\Phi_X\,
  (2\pi)^4\delta^{(4)}(p_B-q-P_X)
  \calT_X(\tau)
  \langle\Bbar|\mathcal J_\gamma^{\dagger\mu}(q)|X\rangle
  \langle X|\mathcal J_\gamma^\nu(q)|\Bbar\rangle.
  \label{eq:fullQCDMeasuredTensor}
\end{align}
To display it in an operator form, introduce a total-momentum operator
$\widehat P_X^\mu$ and an energy-flow measurement operator satisfying
$\widehat{\calT}(\tau)|X\rangle=\calT_X(\tau)|X\rangle$, then
the same tensor in Eq.~\eqref{eq:fullQCDMeasuredTensor} can be written as the cut forward matrix element
\begin{align}
  W_{\BTEC}^{\mu\nu}
  &=\frac1{2m_B}
  \langle\Bbar|
  \mathcal J_\gamma^{\dagger\mu}
  (2\pi)^4\delta^{(4)}(p_B-q-\widehat P_X)
  \widehat{\calT}(\tau)
  \mathcal J_\gamma^\nu
  |\Bbar\rangle_{\rm cut}.
  \label{eq:fullQCDCutOperator}
\end{align}
Factorization can be regarded as the leading-power reduction of the four ingredients in $W_{\BTEC}^{\mu\nu}$.
The current product produces the hard function, the state sum separates into
collinear and soft completeness relations, overall momentum conservation
becomes the jet--soft convolution, and the energy-flow operator reduces to a
collinear measurement at the natural collinear angular scale.

\subsection{Modes and momentum decomposition}

We first identify the leading momentum regions of Eq.~\eqref{eq:fullQCDMeasuredTensor}.
Let $n^\mu=(1,\hat n_J)$ and $\bar n^\mu=(1,-\hat n_J)$, with
$n^2=\bar n^2=0$ and $n\cdot\bar n=2$.
For $\lambda^2\sim\Lambda_{\rm QCD}/Q$, collinear and soft momenta scale as
$  p_n^\mu\sim Q(1,\lambda^2,\lambda)$ and $k_s^\mu\sim Q(\lambda^2,\lambda^2,\lambda^2)$, respectively,
in the $(\bar n\cdot p,n\cdot p,p_\perp)$ components.
The final-state momentum can be separated into modes,
\begin{align}
  P_X^\mu=p_{X_n}^\mu+k_{X_s}^\mu+\cdots.
  \label{eq:modeDecomp}
\end{align}
Note that the total collinear momentum is not the momentum of a single resolved parton.
Its invariant mass is denoted by $s$, i.e., 
\begin{align}
  s=p_{X_n}^2
  =P_-\,n\cdot p_{X_n}+\order(\Lambda_{\rm QCD}^2),
  \label{eq:sdef}
\end{align}
where the leading label constraints
$\bar n\cdot p_{X_n}=P_-$ and vanishing total transverse label have been
used.  Soft momentum cannot change these collinear labels.  Exact momentum
conservation shifts the full large component by $\order(\Lambda_{\rm QCD})$
and leaves a residual total collinear transverse momentum of the same order.
Both effects change Eq.~\eqref{eq:sdef} only by
$\order(\Lambda_{\rm QCD}^2)$.
The small light-cone component fixed by the photon energy contains both
sectors,
\begin{align}
  P_+(E_\gamma)=n\cdot p_{X_n}+\widehat\omega
  +\order\!\left(\frac{\Lambda_{\rm QCD}^2}{Q}\right),
  \label{eq:PplusDecomp}
\end{align}
where $\widehat\omega=n\cdot k_{X_s}\geq0$ is the hadronic soft
light-cone variable entering the hatted shape function, as shown below.
Combining Eqs.~\eqref{eq:sdef} and \eqref{eq:PplusDecomp}, one has
\begin{align}
  s
  &=\big[P_--\bar n\cdot k_{X_s}\big]
  \big[P_+(E_\gamma)-\widehat\omega\big]
  -\boldsymbol k_{X_s\perp}^{\,2}+\cdots
  \nonumber\\
  &=P_-[P_+(E_\gamma)-\widehat\omega]
  +\order(\Lambda_{\rm QCD}^2).
  \label{eq:sargument}
\end{align}
The second line drops both the soft correction to the large component and
the exact transverse-balance term.  Each changes $s$ by
$\order(\Lambda_{\rm QCD}^2)$, or relatively by
$\Lambda_{\rm QCD}/Q$ when $s\sim Q\Lambda_{\rm QCD}$.

\subsection{Radiative current and hard function}

We now turn to the decay matrix element in Eq.~\eqref{eq:fullQCDCurrentDefinition}.
The relevant part of the weak Hamiltonian is~\cite{Buchalla:1995vs}
\begin{align}
  \mathcal H_{\rm eff}^{b\to s\gamma}
  =-\frac{4G_F}{\sqrt2}V_{tb}V_{ts}^*
  \left[C_7^{\rm eff}O_7+C_7'O_7'+\ldots\right],
  \label{eq:Heff}
\end{align}
with
\begin{align}
  O_7&=\frac{e m_b}{16\pi^2}
  \bar s\sigma_{\mu\nu}P_R b\,F^{\mu\nu},\nonumber\\  
  O_7'&=\frac{e m_b}{16\pi^2}
  \bar s\sigma_{\mu\nu}P_L b\,F^{\mu\nu}.
\end{align}
At the hard scale, the direct $O_7$ channel matches onto
\begin{align}
  J_\gamma^{(7)}(0)
  =C_\gamma^{(7)}(m_b,\mu)\,
  \bar\chi_n(0)\Gamma_\gamma^{(7)}h_v(0),
  \label{eq:current}
\end{align}
where $\chi_n=W_n^\dagger\xi_n$ is the gauge-invariant collinear quark field and $h_v$ is the HQET field.
After summing over photon polarizations and neglecting the strange-quark
mass, the two chiralities share the same jet and shape functions, while their
short-distance coefficients enter
\begin{align}
  H_\gamma(m_b,\mu)
  =|C_\gamma^{(7)}|^2+|C_\gamma^{(7')}|^2
  +\order(m_s/m_b).
  \label{eq:Hgamma}
\end{align}
Here $\Gamma_\gamma^{(7,7')}$ denotes the standard leading-power
SCET Dirac structure obtained by matching the electromagnetic dipole
operators $O_7^{(\prime)}$ onto a heavy-to-light current
\cite{Becher:2005fg}. Its explicit form will not be needed below;
all scalar matching factors are absorbed into
$C_\gamma^{(7,7')}$. Suppressing the transverse Lorentz indices and the separate chiral
structures, the polarization-summed current product reduces
schematically to
\begin{align}
  \sum_\lambda\epsilon_\mu^*\epsilon_\nu\,
  \mathcal J_\gamma^{\dagger\mu}\mathcal J_\gamma^\nu
  \longrightarrow
  H_\gamma(m_b,\mu)\,
  \big[\bar h_v\bar\Gamma_\gamma\chi_n\big]
  \big[\bar\chi_n\Gamma_\gamma h_v\big].
  \label{eq:currentProductMatching}
\end{align}
Thus all fluctuations with virtuality of order $m_b^2$ have been absorbed into $H_\gamma$, while the remaining operator contains only collinear and soft fields.
For later use in the decay rate, we factor the dimensionful Born prefactor
from the dimensionless hard function and define
\begin{align}
  \Gamma_\gamma^{(0)}
  =\frac{G_F^2\alpha_{\rm em}m_b^5}{32\pi^4}.
  \label{eq:Gamma0}
\end{align}
At tree level, $\Gamma_\gamma^{(0)}|V_{tb}V_{ts}^*|^2H_\gamma$
reproduces the partonic direct-photon normalization.  With this convention,
the $O_7^{(\prime)}$ Wilson coefficients and their hard radiative corrections
are contained in $H_\gamma$.  The factorization relation below is quoted for
the fully integrated photon direction, for which the angular phase space and
polarization contraction in Eq.~\eqref{eq:fullQCDRateTensor} are included in
this conventional normalization.
Operators other than $O_7$ contribute to the hard coefficient through direct virtual corrections.
When the photon couples to light partons, resolved-photon contributions require
additional soft and jet functions and lie beyond the leading direct-photon
relation considered here.

\subsection{Measurement reduction}

After hard matching, we first ask whether the energy-flow measurement can act
on the collinear state alone.  This question does not involve recoil of the
axis.  In the $B$-meson rest frame,
$\boldsymbol P_X=-\boldsymbol q$, so the tagged photon fixes
$\hat n_J=-\hat n_\gamma$ exactly.  Within a fixed final state, omitting the
soft contribution to the measurement changes neither this axis nor any
collinear particle direction.  Each collinear particle therefore keeps the
same value of $\tau_i$.  This statement does not remove the separate
transverse-momentum correlation between the collinear and soft state sums,
which is analyzed below.

Splitting the final state into collinear and soft particles, while using the
same exact normalization $E_X$ and the same photon-defined axis, gives
\begin{align}
  \calT_X(\tau)
  &=\calT_{X_n}(\tau)+\calT_{X_s}(\tau),
 ~~ \calT_{X_n}(\tau)
  \equiv\frac{1}{E_X}\sum_{i\in X_n}E_i\,
  \delta(\tau-\tau_i),
  ~~
  \calT_{X_s}(\tau)
  \equiv\frac{1}{E_X}\sum_{a\in X_s}E_a\,
  \delta(\tau-\tau_a).
  \label{eq:measurementSectorSplit}
\end{align}
For any bounded angular bin $\mathcal B$, positivity of the energy weights
implies
\begin{align}
  0\leq
  \int_{\mathcal B}\dd\tau\,
  \big[\calT_X(\tau)-\calT_{X_n}(\tau)\big]
  \leq\frac{E_{X_s}}{E_X}
  =\order\!\left(\frac{\Lambda_{\rm QCD}}{Q}\right).
  \label{eq:directSoftMeasurementBound}
\end{align}
Equivalently, the same bound holds against any bounded test function, up to
its supremum norm.  Thus, at leading power,
\begin{align}
  \widehat{\calT}(\tau)
  \doteq\widehat{\calT}_n(\tau)
  +\order\!\left(\frac{\Lambda_{\rm QCD}}{Q}\right),
  \label{eq:measurementReduction}
\end{align}
where $\doteq$ denotes equality in a bounded bin, or equivalently as a
distribution in $\tau$.  Note that no expansion of the angular delta function is
needed: its argument for each collinear particle is unchanged because the
axis is fixed.  The only omitted term is the directly measured soft energy. However,
unbounded positive moments can invalidate this estimate by enhancing
wide-angle soft particles. They will be discussed separately below.

The collinear measurement entering the factorized matrix element also uses
its leading-power energy weight.  Since
$E_i=(\bar n\cdot p_i+n\cdot p_i)/2$ and
$E_X=(Q+P_+)/2$, one obtains
\begin{align}
  \calT_{X_n}^{\rm LP}(\tau)
  &=\sum_{i\in X_n}\frac{\bar n\cdot p_i}{Q}\,
  \delta(\tau-\tau_i),\quad
  \calT_{X_n}(\tau)
  =\calT_{X_n}^{\rm LP}(\tau)
  +\order\!\left(\frac{\Lambda_{\rm QCD}}{Q}\right)
  \label{eq:leadingCollinearMeasurement}
\end{align}
in a bounded bin.  Below, $\calT_{X_n}$ denotes this leading-power form
without an additional superscript.  This convention makes the measured jet
function independent of the external value of $P_+(E_\gamma)$; the physical
$P_+$ dependence enters only through its invariant-mass argument.

\subsection{BPS decoupling and state factorization}

At leading power, soft gluons are removed from the collinear Lagrangian by the BPS field redefinition,
\begin{align}
  \chi_n(x)=Y_n(x)\chi_n^{(0)}(x),
  \qquad
  h_v(x)=Y_v(x)h_v^{(0)}(x).
  \label{eq:BPS}
\end{align}
The current product then contains the soft Wilson-line structures
\begin{align}
  \bar h_v\bar\Gamma_\gamma\chi_n
  &\to
  \bar h_v^{(0)}Y_v^\dagger Y_n
  \bar\Gamma_\gamma\chi_n^{(0)},
~~
  \bar\chi_n\Gamma_\gamma h_v
  \to
  \bar\chi_n^{(0)}\Gamma_\gamma
  Y_n^\dagger Y_vh_v^{(0)}.
  \label{eq:WilsonStructure}
\end{align}
At leading power the initial state has no collinear constituents,
$|\Bbar\rangle=|0\rangle_n\otimes|\Bbar\rangle_s$, while a cut endpoint state
decomposes as $|X\rangle\to|X_n\rangle\otimes|X_s\rangle$.  The completeness
relation in Eq.~\eqref{eq:fullQCDCutOperator} consequently becomes
\begin{align}
  \sum_X\int\dd\Phi_X\,|X\rangle\langle X|
  \;&\longrightarrow
  \left[\sum_{X_n}\int\dd\Phi_{X_n}\,
  |X_n\rangle\langle X_n|\right]
  \otimes
  \left[\sum_{X_s}\int\dd\Phi_{X_s}\,
  |X_s\rangle\langle X_s|\right].
  \label{eq:cutCompletenessFactorization}
\end{align}
Each sum includes the particle multiplicities, species, spins, colors, and
symmetry factors appropriate to its sector.  The overall momentum-conservation
delta function is not part of either completeness relation and continues to
couple the two sectors kinematically.  For example, a BPS-decoupled current
matrix element factorizes as
\begin{align}
  &\langle X_nX_s|
  \bar\chi_n^{(0)}\Gamma_\gamma
  Y_n^\dagger Y_vh_v^{(0)}|\Bbar\rangle
=
  \langle X_n|\bar\chi_n^{(0)}\Gamma_\gamma|0\rangle
  \langle X_s|Y_n^\dagger Y_vh_v^{(0)}|\Bbar\rangle,
  \label{eq:BPSMatrixElementFactorization}
\end{align}
with an analogous relation for the conjugate current.
For the spin-independent measurement considered here, the photon-polarization
sum and the fixed leading-power spin projectors reduce the pair of
$\Gamma_\gamma$ matrices to a kinematics-independent Dirac trace.  We absorb
this constant into the conventional normalization
$\Gamma_\gamma^{(0)}H_\gamma$; the remaining scalar collinear coefficient
defines the measured jet function.  As a result, no explicit $\Gamma_\gamma$ appears
in the final factorization relation.

\subsection{Transverse-momentum reduction}

\begin{figure}[H]
  \centering
  \begin{tikzpicture}[
    x=1cm,y=1cm,>=Latex,
    axis/.style={line width=0.9pt,-{Latex[length=2.2mm]}},
    vec/.style={line width=1.2pt,-{Latex[length=2.5mm]}}
  ]
    \coordinate (O) at (0,0);
    \coordinate (P) at (3.75,0);
    \coordinate (C) at (3.75,-0.78);
    \fill (O) circle (1.2pt) node[below=2pt] {$\Bbar$};
    \draw[axis,dashed,gray!80!black] (O)--(4.25,0)
      node[right,black] {$\hat n_J$};
    \draw[vec,blue!70!black] (O)--(P)
      node[pos=0.62,above=3pt,black] {$P_X$};
    \draw[vec,red!75!black] (O)--(-3.75,0)
      node[left,black] {$q$};
    \draw[vec,black] (O)--(C)
      node[pos=0.60,below=3pt] {$p_{X_n}$};
    \draw[vec,violet!75!black] (C)--(P)
      node[midway,right,black] {$\boldsymbol k_{X_s\perp}$};
  \end{tikzpicture}
  \caption{The tagged photon fixes $P_X$ and $\hat n_J$ exactly.  Separating
  the fixed hadronic momentum into collinear and soft parts requires their
  transverse components to balance.  This state-sum correlation is distinct
  from the measurement reduction.}
  \label{fig:recoilGeometry}
\end{figure}
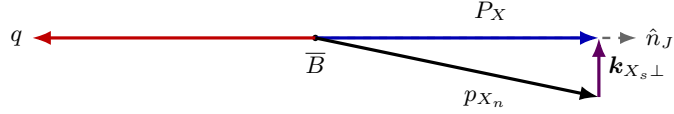

The measurement reduction in Eq.~\eqref{eq:measurementReduction} and the BPS
decoupling solve two different problems: the former removes the directly
measured soft energy, while the latter factorizes the current matrix element.
Neither step removes the exact momentum-conservation delta function.  In
particular, the fixed total momentum is divided between the two sectors as
$
  \boldsymbol p_{X_n\perp}+\boldsymbol k_{X_s\perp}=0
$.
This is a correlation between the collinear and soft state sums.
The geometry is shown in
Fig.~\ref{fig:recoilGeometry}.

We now turn to the four-dimensional delta function in
Eq.~\eqref{eq:fullQCDCutOperator}.  At fixed photon energy,
$n\cdot(p_B-q)=P_+(E_\gamma)\equiv m_B-2E_\gamma$ and
$\bar n\cdot(p_B-q)=Q=m_B$.  Using
$\delta^{(4)}(l)=2\delta(n\cdot l)\delta(\bar n\cdot l)
\delta^{(2)}(l_\perp)$, one finds
\begin{align}
  & \delta^{(4)}
  (p_B-q-p_{X_n}-k_{X_s})
=2 
  \delta\!\left(Q-\bar n\cdot p_{X_n}-\bar n\cdot k_{X_s}\right)
  \delta\!\left(P_+(E_\gamma)-n\cdot p_{X_n}-n\cdot k_{X_s}\right)
  \delta^{(2)}\!\left(p_{X_n\perp}+k_{X_s\perp}\right).
  \label{eq:fourDeltaLightCone}
\end{align}
The second delta function is the small-component part of overall momentum
conservation;
the large component presents no new complication.  Since
$\bar n\cdot k_{X_s}\sim\Lambda_{\rm QCD}$ while
$\bar n\cdot p_{X_n}\sim Q$, it fixes
$\bar n\cdot p_{X_n}=Q$ at leading power, with a relative correction of order
$\Lambda_{\rm QCD}/Q$.

The transverse component requires one additional treatment.  A generic
collinear constituent and the total soft state have transverse momenta
\begin{align}
  p_{i\perp}\sim Q\lambda\sim\sqrt{Q\Lambda_{\rm QCD}},
  \qquad
  k_{X_s\perp}\sim Q\lambda^2\sim\Lambda_{\rm QCD},
  \qquad
  \lambda^2\sim\frac{\Lambda_{\rm QCD}}{Q}.
  \label{eq:SCETTransverseScales}
\end{align}
It is important to distinguish $p_{i\perp}$ in
Eq.~\eqref{eq:SCETTransverseScales} from the total collinear momentum
$p_{X_n\perp}$.  The former is the transverse momentum of
an individual collinear constituent relative to the jet direction and hence
resolves the internal collinear structure.  By contrast,
$p_{X_n\perp}=\sum_{i\in X_n}p_{i\perp}$ is the total transverse momentum of
the collinear state.  The constituent momenta of order
$\sqrt{Q\Lambda_{\rm QCD}}$ can cancel in this vector sum, while the exact
transverse constraint in Eq.~\eqref{eq:fourDeltaLightCone} fixes
$p_{X_n\perp}=-k_{X_s\perp}\sim\Lambda_{\rm QCD}$.  Therefore, the effect of
soft recoil cannot be estimated by directly comparing $k_{X_s\perp}$ with
the constituent momentum $p_{i\perp}$. It must be examined after the
transverse convolution in the measured state sum.

After the transverse delta function is integrated, the measured collinear
state sum is evaluated at $p_{X_n\perp}=-k_{X_s\perp}$.  The leading-power
approximation amounts to replacing this value by the one at
$p_{X_n\perp}=0$.  To estimate the difference, one expands the state sum in
its total transverse-momentum argument.  For a spinless $B$ meson and a BTEC
averaged over the azimuth about the photon axis, no transverse vector is
available to form a scalar linear in $p_{X_n\perp}$.  Hence the linear term
vanishes, and the first recoil correction is proportional to
$\boldsymbol k_{X_s\perp}^{\,2}$.

For a smooth spectrum or a finite angular bin, two collinear scales control
the coefficient of this quadratic correction.  (1) The angular
measurement resolves $p_{i\perp}^2\sim\tau$. For
$E_i=z_iQ/2$ with $z_i=\order(1)$, Eq.~\eqref{eq:tau} gives
$|p_{i\perp}|\simeq E_i\theta_{iJ}\sim z_i\sqrt{\tau}$.  A collective
transverse displacement by $k_{X_s\perp}$ therefore produces a relative
correction of order $\boldsymbol k_{X_s\perp}^{\,2}/\tau$.  (2) Exact
transverse balance shifts the collinear invariant mass by
$\order(\boldsymbol k_{X_s\perp}^{\,2})$, as shown in
Eq.~\eqref{eq:sargument}, and hence produces a relative correction of order
$\boldsymbol k_{X_s\perp}^{\,2}/s$.  Denoting the combined soft transverse
recoil correction by $\Delta_\perp W_{\BTEC}$, one obtains
\begin{align}
  \frac{\Delta_\perp W_{\BTEC}}{W_{\BTEC}}
  &\sim
  \underbrace{\mathcal{O}\left(\frac{\boldsymbol k_{X_s\perp}^{\,2}}{\tau}\right)}
  _{\text{angular migration}}
  +
  \underbrace{\mathcal{O}\left(\frac{\boldsymbol k_{X_s\perp}^{\,2}}{s}\right)}
  _{\text{invariant-mass shift}}
  \nonumber\\
  &\sim
  \mathcal{O}\left(\frac{\Lambda_{\rm QCD}^2}{\tau}\right)
  + \mathcal{O}\left(\frac{\Lambda_{\rm QCD}^2}{s}\right)
  \sim \mathcal{O}\left(\frac{\Lambda_{\rm QCD}}{Q}\right),
  \qquad \tau\sim s\sim Q\Lambda_{\rm QCD}.
  \label{eq:transverseBalanceCorrection}
\end{align}
Thus, at the natural angular scale, the soft transverse recoil is power
suppressed.  Inside the physical rate one may therefore use the
leading-power shorthand
\begin{align}
  \delta^{(d-2)}\!\left(
  p_{X_n\perp}+k_{X_s\perp}\right)
  \xrightarrow[\text{ transverse convolution}]{\rm LP}
  \delta^{(d-2)}\!\left(p_{X_n\perp}\right).
  \label{eq:transverseDeltaReduction}
\end{align}
Appendix~\ref{app:transverse}
extends the argument at the function-level and the small-angle limit.

By contrast,
$n\cdot p_{X_n}\sim n\cdot k_{X_s}\sim\Lambda_{\rm QCD}$, so the remaining
small-component constraint cannot be expanded.  It is separated through
\begin{align}
  &\delta(P_+(E_\gamma)-n\cdot p_{X_n}-n\cdot k_{X_s})
=
  \int_0^\infty\dd\widehat\omega\,
  \delta(P_+(E_\gamma)-n\cdot p_{X_n}-\widehat\omega)\,
  \delta(\widehat\omega-n\cdot k_{X_s}).
  \label{eq:hatOmegaInsert}
\end{align}
Before the soft-state sum is carried out, its cut factor is
\begin{align}
  &\sum_{X_s}\int\dd\Phi_{X_s}\,
  \langle\Bbar|
  \bar h_v^{(0)}Y_v^\dagger Y_n|X_s\rangle
  \langle X_s|Y_n^\dagger Y_vh_v^{(0)}|\Bbar\rangle
  \delta(\widehat\omega-n\cdot k_{X_s})
 =
  \langle\Bbar|
  \bar h_v^{(0)}Y_v^\dagger Y_n
  \delta(\widehat\omega-\bar\Lambda+i n\cdot\partial)
  Y_n^\dagger Y_vh_v^{(0)}|\Bbar\rangle.
  \label{eq:softCompletenessReduction}
\end{align}
The equality follows from soft completeness and translation invariance.
Here $\bar\Lambda=m_B-m_b$ in the residual-mass convention used to define
$h_v$.  Acting on the soft matrix element, $i n\cdot\partial$ measures the
unhatted residual variable
$\ell=\bar\Lambda-n\cdot k_{X_s}=\bar\Lambda-\widehat\omega$.
Thus the explicit $X_s$ sum has become the soft forward matrix element rather
than being omitted.

Combining Eqs.~\eqref{eq:measurementReduction},
\eqref{eq:BPSMatrixElementFactorization},
\eqref{eq:transverseDeltaReduction}, \eqref{eq:hatOmegaInsert}, and
\eqref{eq:softCompletenessReduction} gives the factorized representation of
$W_{\BTEC}$.  From this point on, $\sum_{X_n}$ includes the collinear
phase-space integral specified above.  After this spin projection, one
obtains the following expression, where $P_+$ is shorthand for the derived
quantity $P_+(E_\gamma)$ at fixed photon energy and is not an independent
spectrum variable:
\begin{align}
  W_{\BTEC}(\tau;q)
  &\longrightarrow
  H_\gamma\!\int_0^\infty\!\dd\widehat\omega\,
  \underbrace{\sum_{X_n}\!(2\pi)^{d-1}
  \delta\!\left(Q-\bar n\cdot p_{X_n}\right)
  \delta^{(d-2)}\!\left(p_{X_n\perp}\right)
  \langle0|\chi_n^{(0)}|X_n\rangle
  \langle X_n|\bar\chi_n^{(0)}|0\rangle
  \delta(P_+-n\cdot p_{X_n}-\widehat\omega)
  \calT_{X_n}(\tau)}_{\text{measured collinear cut matrix element}}
  \nonumber\\
  &\qquad\times
  \underbrace{\frac1{2m_B}
  \langle\Bbar|
  \bar h_v^{(0)}Y_v^\dagger Y_n
  \delta(\widehat\omega-\bar\Lambda+i n\cdot\partial)
  Y_n^\dagger Y_vh_v^{(0)}
  |\Bbar\rangle}_{\text{soft matrix element}}
  +\text{power corrections}.
  \label{eq:operatorLevelReduction}
\end{align}
The two underbraced factors become, with the conventional normalizations
restored, the measured jet function and the leading shape function.  At this
stage the collinear factor still contains the physical small-component
constraint
$\delta(P_+-n\cdot p_{X_n}-\widehat\omega)$ inherited from
Eq.~\eqref{eq:hatOmegaInsert}.  For a specified soft momentum
$\widehat\omega$, this
delta function fixes $n\cdot p_{X_n}$.  Below we convert it into the
invariant-mass projector used in the standard $s$-space jet function.

The soft factor in Eq.~\eqref{eq:operatorLevelReduction} is the leading shape
function~\cite{Neubert:1993um,Bosch:2004th}
\begin{align}
  \widehat S_B(\widehat\omega,\mu)
  &=\frac{1}{2m_B}
  \langle\Bbar(v)|
  \bar h_v^{(0)}Y_v^\dagger Y_n
  \delta(\widehat\omega-\bar\Lambda+i n\cdot\partial)
  Y_n^\dagger Y_vh_v^{(0)}
  |\Bbar(v)\rangle.
  \label{eq:shapeWilson}
\end{align}
It is equivalent to the gauge-covariant definition
\begin{align}
  \widehat S_B(\widehat\omega,\mu)
  =\frac{1}{2m_B}
  \langle\Bbar(v)|\bar h_v
  \delta(\widehat\omega-\bar\Lambda+i n\cdot D)h_v|\Bbar(v)\rangle.
\end{align}
Equivalently, in terms of the conventional unhatted variable $\ell$,
\begin{align}
\begin{aligned}
  \widehat\omega&=\bar\Lambda-\ell,\quad
  \widehat S_B(\widehat\omega,\mu)
  =S_B(\bar\Lambda-\widehat\omega,\mu),\quad
  -\infty<\ell\leq\bar\Lambda
  \Longleftrightarrow
  0\leq\widehat\omega<\infty ,
\end{aligned}
\label{eq:shapeSupportMap}
\end{align}
where the last one is the intrinsic support of the shape function.  In
particular, its renormalized radiative tail extends to
$\widehat\omega\to\infty$.
Changing the heavy-quark mass convention shifts $\ell$ and $\bar\Lambda$
together and leaves the hadronic variable $\widehat\omega$ unchanged.
The heavy quark and the soft spectator are contained in this matrix element; at the natural collinear angular scale, no leading BTEC insertion acts on the soft final state.

\subsection{Collinear matrix element and factorization relation}

To separate the collinear state sum from the soft variable
$\widehat\omega$, we first
define it for an independent collinear invariant mass $s$.  The measured jet
function is normalized in $d$ dimensions as
\begin{align}
  J_{\rm TEC}^{q}(s,\tau,\mu)
  &\equiv\mathcal N_J\sum_{X_n}
  {\rm Tr}\!\left[
  \frac{\slashed{\bar n}}{2}
  \langle0|\chi_n^{(0)}(0)|X_n\rangle
  \langle X_n|\bar\chi_n^{(0)}(0)|0\rangle
  \right]
  \nonumber\\
  &\quad\times(2\pi)^{d-1}
  \delta(Q-\bar n\cdot p_{X_n})
  \delta^{(d-2)}(p_{X_n\perp})
  \delta(s-p_{X_n}^2)
  \calT_{X_n}(\tau).
  \label{eq:Jdefinition}
\end{align}
Here the color trace is normalized by $1/N_c$ and $\mathcal N_J=1$ in our
convention.  The cut-state measure, momentum-projector Jacobian, and
tree-level normalization are detailed in Appendix~\ref{app:norm}.  The first
two delta functions are the same large- and
transverse-momentum constraints that appear in
Eq.~\eqref{eq:operatorLevelReduction}.  In particular,
$\delta^{(d-2)}(p_{X_n\perp})$ fixes the total collinear transverse momentum
relative to the fixed photon axis after the leading-power transverse-momentum
reduction.  The third delta function does not impose an
additional constraint on Eq.~\eqref{eq:operatorLevelReduction}.  Instead, it
defines the invariant-mass argument of the jet function through
$s=p_{X_n}^2$, before the physical value of $s$ is selected by the
small-component delta function in Eq.~\eqref{eq:operatorLevelReduction}.
To relate the two representations, note that the large- and
transverse-momentum constraints imply
$p_{X_n}^2=P_-\,n\cdot p_{X_n}+\order(\Lambda_{\rm QCD}^2)$.
The small-component delta function in Eq.~\eqref{eq:operatorLevelReduction}
can therefore be rewritten as
\begin{align}
  \delta(P_+(E_\gamma)-n\cdot p_{X_n}-\widehat\omega)
  =P_-\delta\!\left(P_-[P_+(E_\gamma)-\widehat\omega]
  -p_{X_n}^2\right)
  +\text{power corrections}.
  \label{eq:deltaConvert}
\end{align}
The factor $P_-$ is the Jacobian of the change from the small light-cone
momentum to the invariant mass.  The delta function on the right-hand side
has precisely the form of the invariant-mass projector in
Eq.~\eqref{eq:Jdefinition}, evaluated at
$s=P_-[P_+(E_\gamma)-\widehat\omega]$.  Thus
Eq.~\eqref{eq:deltaConvert} converts the
collinear factor in Eq.~\eqref{eq:operatorLevelReduction} into
$J_{\rm TEC}^q(P_-[P_+(E_\gamma)-\widehat\omega],\tau,\mu)$.

Combining this result
with the full-QCD rate in Eq.~\eqref{eq:fullQCDRateTensor}, the hard matching
in Eq.~\eqref{eq:currentProductMatching}, the normalization in
Eq.~\eqref{eq:Gamma0}, and the soft matrix element gives the leading BTEC
factorization relation:
\begin{align}
  \frac{\dd^2\Gamma_{\BTEC}^{s\gamma}}
       {\dd E_\gamma\,\dd\tau}
  &=2\Gamma_\gamma^{(0)}|V_{tb}V_{ts}^*|^2
  H_\gamma(m_b,\mu)P_-
  \int_0^{P_+(E_\gamma)}\dd\widehat\omega\,
  J_{\rm TEC}^{q}\!\left(
  P_-[P_+(E_\gamma)-\widehat\omega],\tau,\mu\right)
  \widehat S_B(\widehat\omega,\mu)
  \nonumber\\
  &\quad+\text{power corrections},
  \label{eq:factorization}
\end{align}
where the overall factor $2=|\dd P_+/\dd E_\gamma|$ converts the conventional
$P_+$ spectrum normalization to the photon-energy spectrum.  In particular,
it ensures that the tree-level endpoint term integrates to the Born width.

Eq.~\eqref{eq:factorization} is the main result of this work.  It applies
at the natural collinear angular scale; the omitted terms include
relative corrections of order $\Lambda_{\rm QCD}/Q$, $s/Q^2$, $\tau/Q^2$,
and $\Lambda_{\rm QCD}^2/\tau$.  The last term is specific to the angular
sensitivity to soft--collinear transverse-momentum balance; the accompanying
invariant-mass shift is $\Lambda_{\rm QCD}^2/s$ and is already of order
$\Lambda_{\rm QCD}/Q$ in the endpoint region.
The intrinsic support $\widehat\omega\geq0$ in
Eq.~\eqref{eq:shapeSupportMap}, together with the jet-function support
$s\geq0$, gives the physical convolution range
$0\leq\widehat\omega\leq P_+(E_\gamma)$ displayed explicitly in
Eq.~\eqref{eq:factorization}.

The factorization relation separates the perturbatively calculable and
nonperturbative ingredients by their characteristic scales.  The hard
function $H_\gamma$ contains fluctuations at
$\mu_H\sim m_b$ and can be calculated order by order in
$\alpha_s(\mu_H)$.  In the natural collinear region,
$\mu_J^2\sim s\sim\tau\sim Q\Lambda_{\rm QCD}$, the measured jet function
$J_{\rm TEC}^q$ is also perturbatively calculable.  By contrast, the shape
function $\widehat S_B$ probes soft momenta
$\widehat\omega\sim\Lambda_{\rm QCD}$ and cannot be determined by
perturbation theory in this region. At leading power
it is the same universal shape function that appears in the ordinary photon
spectrum.  Its renormalization-group evolution and
perturbative tail are calculable, while its soft functional form must be
extracted from data, or lattice simulations~\cite{Wang:2025uap,Xiong:2026fvs}.
  A complete treatment of subleading-power and resolved-photon
contributions generally introduces additional nonperturbative functions and
is outside the present work.

Therefore, the only new perturbative ingredient required for the angular
spectrum is $J_{\rm TEC}^q$.  In the next section, we calculate this function
at one loop, determine its angular dependence, and examine its ultraviolet
renormalization.  Before carrying out the calculation, its definition gives
the inclusive consistency condition
\begin{align}
  \int_0^\infty\dd\tau\,
  J_{\rm TEC}^{q}(s,\tau,\mu)\overset{\rm LP}{=}J_q(s,\mu),
  \label{eq:inclusiveSum}
\end{align}
where $J_q$ is the standard inclusive quark jet function
~\cite{Becher:2006qw,Bruser:2018rad}.
This relation follows because Eq.~\eqref{eq:leadingCollinearMeasurement} and the large-component
projector give
$\int\dd\tau\,\calT_{X_n}=\sum_i\bar n\cdot p_i/Q=1$.
Eq.~\eqref{eq:inclusiveSum} therefore recovers the ordinary photon-energy
factorization theorem at leading power.
Note that Eq.~\eqref{eq:inclusiveSum} is a statement about the zeroth angular moment only. It does not imply that the collinear factorization theorem is valid pointwise at $\tau\sim Q^2$. Wide-angle soft radiation contributes only at relative order $\Lambda_{\rm QCD}/Q$ to the unweighted integral, but the additional factor of $\tau\sim Q^2$ promotes it to order $Q\Lambda_{\rm QCD}$ in the first moment, the same order as the collinear contribution. Positive angular moments therefore cannot, in general, be obtained from the leading-power collinear jet function alone.

\section{One-loop measured quark jet function}
\label{sec:jet}

As noted above, the measured jet function can be calculated with perturbation theory in the hard-collinear region. We calculate Eq.~\eqref{eq:Jdefinition} in dimensional regularization with
$d=4-2\epsilon$ and the $\overline{\rm MS}$ scheme.
At tree level, the cut state consists of one quark aligned with the recoil axis, one has
\begin{align}
  J_{\rm TEC}^{q(0)}(s,\tau)=\delta(s)\delta(\tau).
  \label{eq:Jtree}
\end{align}

\subsection{Real emission}

At next-to-leading order (NLO), the real cut is
$q^*\to q(p_1)+g(p_2)$.
Let the gluon carry the large light-cone fraction
$x=\bar n\cdot p_2/Q$, so the quark carries $1-x$.
In the frame with vanishing total transverse momentum, one has
\begin{align}
  &\bar n\cdot p_1=(1-x)Q, ~~~~p_{1\perp}=k_\perp,
  \nonumber\\
&  \bar n\cdot p_2=xQ,~~~~p_{2\perp}=-k_\perp.
\end{align}
Using the on-shell conditions, the final-state invariant mass is
\begin{align}
  s=(p_1+p_2)^2=\frac{k_\perp^2}{x(1-x)}.
  \label{eq:sx}
\end{align}
The two angular arguments are $  \tau_q=s\frac{x}{1-x},
  \tau_g=s\frac{1-x}{x}$ for quark and gluon, respectively.
With $E_i=(\bar n\cdot p_i+n\cdot p_i)/2$ and
$E_{qg}=(Q+s/Q)/2$, the on-shell relations imply
$E_q/E_{qg}=1-x+\order(s/Q^2)$ and
$E_g/E_{qg}=x+\order(s/Q^2)$.  The leading-power measurement on the
two-parton state is therefore
\begin{align}
\calT_{qg}(s,x;\tau)
  =(1-x)\delta\!\left(\tau-s\frac{ x}{1-x}\right)
  +x\delta\!\left(\tau-s\frac{1-x}{x}\right).
  \label{eq:Tqg}
\end{align}
On can immediately check that the first moment obeys the event-level identity at the two-parton level,
\begin{align}
  \int_0^\infty\dd\tau\,\tau\,\calT_{qg}(s,x;\tau)
  =(1-x)\tau_q+x\tau_g=s.
  \label{eq:partonicFirstMoment}
\end{align}
This relation is a nontrivial kinematic check on the real-emission measurement.
At fixed $s$, however, the limits $x\to0,1$ send one of the measured angles outside the collinear region.
Consequently, taking the unbounded first moment after the collinear phase-space expansion requires the physical angular boundary and wide-angle matching.

\begin{figure}[t]
  \centering
  \includegraphics[width=0.60\linewidth]{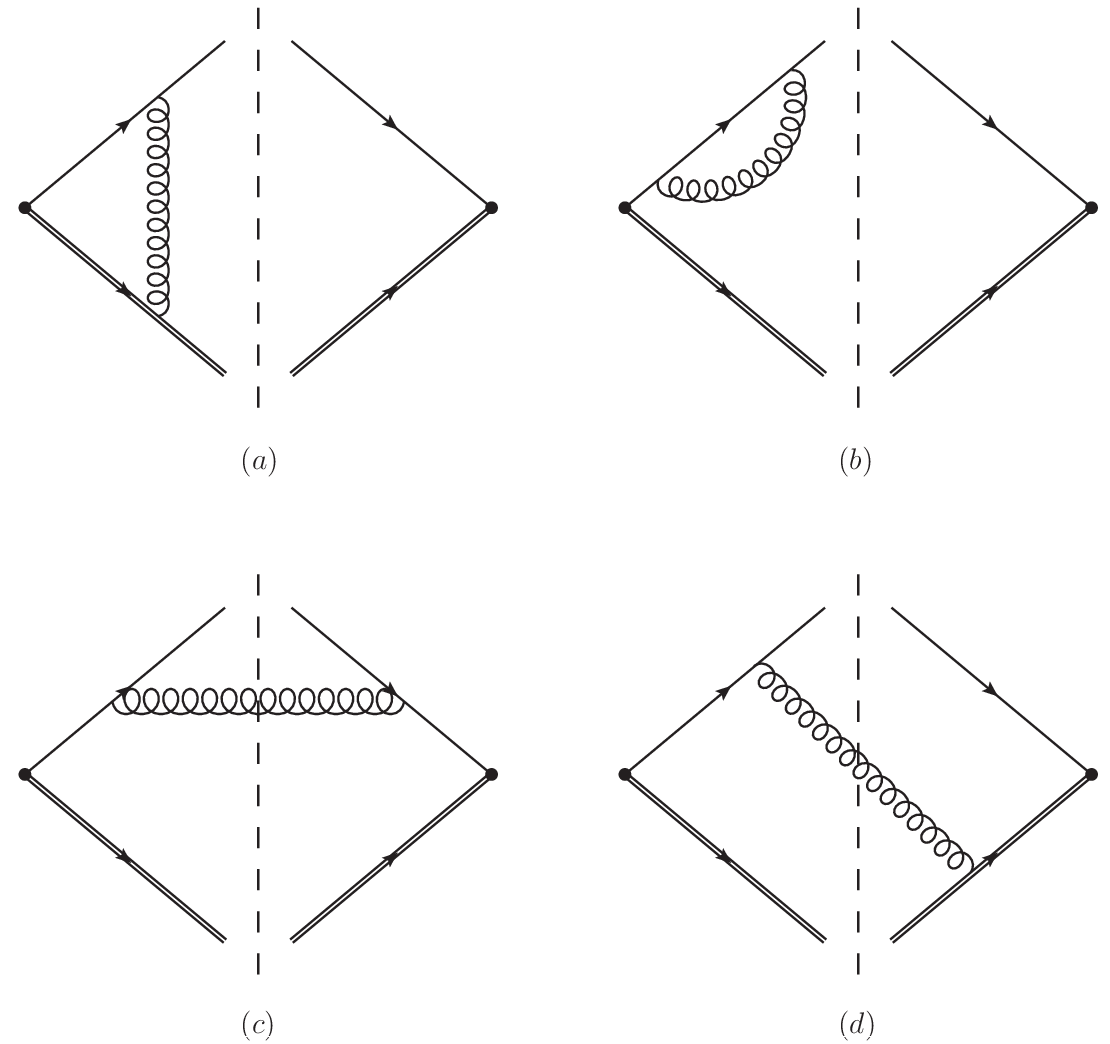}
  \caption{One-loop cut diagrams for the measured quark jet function.
  The dashed vertical line denotes the final-state cut. Panels (a) and (b) represent the virtual corrections, whereas (c) and (d) are the real corrections.
  The real contribution contains the quark-emission square and its interference with emission from the Wilson line in $\chi_n$; the virtual contribution is local in $s$ and $\tau$. The conjugate diagrams are not presented.}
  \label{fig:jetgraphs}
\end{figure}

The relevant cut graphs are shown in Fig.~\ref{fig:jetgraphs}.
The gauge-invariant collinear building block combines emission from the quark line with emission from the collinear Wilson line.

The invariant-mass projector in Eq.~\eqref{eq:Jdefinition} fixes
$k_\perp^2=sx(1-x)$, so the magnitude of $k_\perp$ is no longer an
independent variable at fixed $s$.  After using this relation in the
two-body cut phase space, we define the fixed-$s$ real-cut density
$\mathcal W_{qg}(s,x;\epsilon)$.  Its relation to the measured real jet
function is
\begin{align}
  J_{\rm TEC}^{q,R}(s,\tau)
  &=\int_0^1\dd x\,
  \mathcal W_{qg}(s,x;\epsilon)\,
  \calT_{qg}(s,x;\tau),	\label{eq:WJ}
\end{align}
where $\mathcal{W}_{qg}$ can be extracted by calculating the diagrams in Fig.~\ref{fig:jetgraphs}. The result of the real emission from Fig.~\ref{fig:jetgraphs}(c)(d) reads
\begin{align}
	  \mathcal W_{qg}(s,x;\epsilon)
	&=\frac{\alpha_s C_F}{2\pi}
	\frac{e^{\gamma_E\epsilon}}{\Gamma(1-\epsilon)}
	\left(\frac{\mu^2}{s}\right)^\epsilon
	\frac{1}{s}[x(1-x)]^{-\epsilon}
	\left[\frac{1+(1-x)^2}{x}-\epsilon x\right].
	\label{eq:Wqg}
\end{align}
Substituting Eq.~\eqref{eq:Tqg} into Eqs.~\eqref{eq:WJ} and
\eqref{eq:Wqg}, then performing
the $x$ integral for each of the two measurement terms gives, for
$s,\tau>0$,
\begin{align}
  J_{\rm TEC}^{q,R}(s,\tau)
  =\frac{\alpha_s C_F}{2\pi}
  \frac{e^{\gamma_E\epsilon}}{\Gamma(1-\epsilon)}
  \frac{\mu^{2\epsilon}}{s^{2+\epsilon}}
  g_\epsilon\!\left(\frac{\tau}{s}\right),
  \label{eq:JrealEps}
\end{align}
where
\begin{align}
  g_\epsilon(r)
  =\frac{r^{-\epsilon}}{(1+r)^{4-2\epsilon}}
  \left[\frac{2}{r}+3+3r+2r^2-\epsilon(1+r)\right].
  \label{eq:geps}
\end{align}
In four dimensions,
\begin{align}
  g(r)=\frac{2/r+3+3r+2r^2}{(1+r)^4}.
  \label{eq:g}
\end{align}
The two asymptotic limits of $g(r)$ are
\begin{align}
  g(r)&=\frac2r-5+\order(r),\qquad r\to0,
  \nonumber\\
  g(r)&=\frac2{r^2}-\frac5{r^3}+\order(r^{-4}),
  \qquad r\to\infty.
  \label{eq:gasymptotics}
\end{align}
The small-$r$ singularity is the soft-collinear limit in which the gluon becomes soft while the quark remains near the tagged axis.
The large-$r$ behavior also verifies the support statement following
Eq.~\eqref{eq:inclusiveSum}.  Although the exact event-level variable obeys
$0\leq\tau\leq Q^2$, the collinear two-parton measurement extends to
infinite $\tau$ when a parton approaches an energy endpoint.  Since
$g(r)\sim2/r^2$, the contribution from $\tau>Q^2$ scales as $1/Q^2$, which
is suppressed by $s/Q^2$ relative to the natural inclusive scaling $1/s$.
The omitted wide-angle soft energy is likewise suppressed by
$\Lambda_{\rm QCD}/Q$ in the inclusive normalization,  thus the auxiliary
upper limit in Eq.~\eqref{eq:inclusiveSum} has only power-suppressed
consequences for the physical inclusive rate.

\subsection{Jet function as a two-dimensional distribution}

The singularities of $J_{\rm TEC}^{q,R}(s,\tau)$ at $s=0$ and $\tau=0$ require a prescription. Before converting the real-emission result into a two-dimensional
distribution, we explain why no separate virtual contribution appears below.
For the one-particle cut presented in Fig.~\ref{fig:jetgraphs}(a)(b), the invariant-mass projector enforces $s=p^2=0$ and
the measurement reduces to its tree-level value, so a virtual correction can
only be proportional to $\delta(s)\delta(\tau)$.  The collinear quark is
massless and on shell, and dimensional regularization is used for both
ultraviolet and infrared singularities.  The one-loop quark and Wilson-line
virtual integrals are therefore scaleless and vanish in pure
dimensional regularization, i.e., 
\begin{align}
  J_{\rm TEC}^{q,V,{\rm bare}}(s,\tau)
  =0\times\delta(s)\delta(\tau)=0.
  \label{eq:JvirtualScaleless}
\end{align}
This vanishing result represents the cancellation of ultraviolet and infrared
poles in scaleless integrals, but it does not imply the absence of ultraviolet
renormalization.  The poles exposed by expanding the real-emission result as a
distribution determine the usual quark-jet counterterm.

We now split the one-loop real emission Eq.~\eqref{eq:JrealEps} into a regularized part and a delta function part. We first define the angular plus distribution on the auxiliary half-line
$\tau\geq0$.  For fixed $s>0$ and test function
$f\in C_c^\infty([0,\infty))$, its action is
\begin{align}
  &\int_0^\infty\dd\tau
  \left[\frac1{s^2}g\!\left(\frac\tau s\right)\right]_{\tau+}
  f(\tau)
  =\frac1s\int_0^\infty\dd r\,
  g(r)[f(sr)-f(0)].
  \label{eq:tauplus}
\end{align}
Since $g(r)=\order(r^{-2})$ at large $r$, this action also extends to the
constant function $f=1$ and gives zero.  This is the half-line normalization
used in the inclusive sum rule.  If the angular integration is instead
restricted to a finite upper limit $\tau_{\max}$, the same half-line
distribution gives
\begin{align}
  &\int_0^{\tau_{\max}}\dd\tau
  \left[\frac1{s^2}g\!\left(\frac\tau s\right)\right]_{\tau+}
  f(\tau)
 =
  \frac1s\int_0^{R_{\max}}\dd r\,g(r)
  [f(sr)-f(0)]
  -\frac{f(0)}s\int_{R_{\max}}^\infty\dd r\,g(r),
  \label{eq:tauplusFinite}
\end{align}
where $R_{\max}={\tau_{\max}}/s$. Thus a finite angular cutoff does not define a different plus prescription.
It restricts the half-line distribution and retains its large-$\tau$ tail
through the last term in Eq.~\eqref{eq:tauplusFinite}.

Because the jet function is a distribution in both $s$ and $\tau$, however, the fixed-$s$ statement is not sufficient by itself.
Let the test function $\varphi\in C_c^\infty([0,\infty)^2)$,
the two-dimensional distribution inherited from the real-emission integral acts as
\begin{align}
  \left\langle \mathcal K,\varphi\right\rangle
  &\equiv
  \int_0^\infty\frac{\dd s}{s}
  \int_0^\infty\dd r\,g(r)
  \left[\varphi(s,sr)-\varphi(s,0)\right],
  \nonumber\\
  \mathcal K(s,\tau)
  &\equiv
  \left[\frac1{s^2}g\!\left(\frac\tau s\right)\right]_{\tau+}.
  \label{eq:twoDaction}
\end{align}
We now show that this action is finite at every boundary of the quadrant, therefore the prescription for the superficial singularity $s=0$ is not necessary.
As $r\to0$, $g(r)=\order(1/r)$, while the bracket is $\order(sr)$, so the angular endpoint is integrable.
For $r\to\infty$, $g(r)=\order(1/r^2)$.
More explicitly, let
$M_0=\|\varphi\|_\infty$ and
$M_1=\|\partial_\tau\varphi\|_\infty$.
The mean-value theorem and boundedness give
\begin{align}
  |\varphi(s,sr)-\varphi(s,0)|
  \leq\min(M_1sr,2M_0)
  =C_\varphi\min\!\left(\frac{sr}{S_\varphi},1\right),
  \label{eq:testFunctionBound}
\end{align}
where $C_\varphi=2M_0$ and $S_\varphi=2M_0/M_1$ are fixed by the test
function.  Splitting the $r$ integral at $r=1$ and
$r=S_\varphi/s$ therefore yields, for $s\ll S_\varphi$,
\begin{align}
  \left|
  \frac1s\int_0^\infty\dd r\,g(r)
  [\varphi(s,sr)-\varphi(s,0)]
  \right|
  \leq C_\varphi'
  \left[1+\ln\!\left(\frac{S_\varphi}{s}\right)\right],
  \label{eq:twoDBound}
\end{align}
where $C_\varphi'$ is a constant. This logarithmic bound is locally integrable at $s=0$.
Thus Eq.~\eqref{eq:twoDaction} already defines a genuine distribution in $(s,\tau)$.
No joint-plus prescription and no additional subtraction at $s=0$ are required at this order.
All distributions singular at the $s$ endpoint are contained in the inclusive jet function multiplying $\delta(\tau)$ and are fixed by the inclusive sum rule.
Note that the compact-support condition is essential; the unbounded test function $\varphi\propto\tau$ probes
$r g(r)\sim2/r$ at large $r$ and is not in the domain of the collinear distribution.
This logarithmic sensitivity signals the missing wide-angle region, not a need for another endpoint plus prescription.

We then turn to the delta function part. The coefficient of $\delta(\tau)$ can be obtained directly rather than fixed only by the final sum rule.
Integrating Eq.~\eqref{eq:JrealEps} over the angular variable gives
\begin{align}
  \int_0^\infty\dd\tau\,
  J_{\rm TEC}^{q,R}(s,\tau)
  &=\frac{\alpha_s C_F}{2\pi}
  \frac{\mu^{2\epsilon}}{s^{1+\epsilon}}
  \frac{e^{\gamma_E\epsilon}}{\Gamma(1-\epsilon)}
  \int_0^\infty\dd r\,g_\epsilon(r) \nonumber\\
  &=\frac{\alpha_s C_F}{2\pi}
  \frac{\mu^{2\epsilon}}{s^{1+\epsilon}}
  \frac{e^{\gamma_E\epsilon}}{\Gamma(1-\epsilon)}
  \frac{(-4+\epsilon)\Gamma^2(2-\epsilon)}
  {\epsilon\Gamma(3-2\epsilon)}.
  \label{eq:inclusiveRealProjection}
\end{align}
To expand the endpoint singularity, we use the dimensionful star
distributions which is standard in inclusive $B$-decay factorization~\cite{Bosch:2004th,Becher:2006qw}.  For a smooth test function $T(s)$,
an arbitrary upper limit $z>0$, and an independent subtraction scale $u>0$,
they are defined by
\begin{align}
\begin{aligned}
  \int_0^z\dd s\,T(s)
  \left(\frac1s\right)_{\!*}^{[u]}
  &=
  \int_0^z\dd s\,\frac{T(s)-T(0)}s
  +T(0)\ln\frac zu,
  \\
  \int_0^z\dd s\,T(s)
  \left(\frac{\ln(s/u)}s\right)_{\!*}^{[u]}
  &=
  \int_0^z\dd s\,\frac{T(s)-T(0)}s\ln\frac su
  +\frac{T(0)}2\ln^2\frac zu.
\end{aligned}
\label{eq:sStarDefinition}
\end{align}
The subtraction scale $u$ is independent of the renormalization scale $\mu$
at the level of the definition.  The endpoint expansion is
\begin{align}
\begin{aligned}
  \frac{\mu^{2\epsilon}\theta(s)}{s^{1+\epsilon}}
  &=
  \left(\frac{\mu^2}{u}\right)^\epsilon
  \left[
  -\frac1\epsilon\delta(s)
  +\left(\frac1s\right)_{\!*}^{[u]}
  -\epsilon
  \left(\frac{\ln(s/u)}s\right)_{\!*}^{[u]}
  +\order(\epsilon^2)
  \right]
  \\
  &\xrightarrow{\,u=\mu^2\,}
  -\frac1\epsilon\delta(s)
  +\left(\frac1s\right)_{\!*}^{[\mu^2]}
  -\epsilon
  \left(\frac{\ln(s/\mu^2)}s\right)_{\!*}^{[\mu^2]}
  +\order(\epsilon^2).
\end{aligned}
  \label{eq:sEndpointExpansion}
\end{align}
The hatted shape function has intrinsic support
$0\leq\widehat\omega<\infty$, while the inclusive jet function has support
$s\geq0$.  In the physical convolution these conditions combine to give
$0\leq\widehat\omega\leq P_+$ and hence
$0\leq s=Q(P_+-\widehat\omega)\leq QP_+$.
This finite convolution interval supplies the upper limit $z$ in
Eq.~\eqref{eq:sStarDefinition}. For simplicity, we choose $u=\mu^2$ in the jet
function below.
Combining Eqs.~\eqref{eq:inclusiveRealProjection} and
\eqref{eq:sEndpointExpansion},
the
inclusive real projection becomes
\begin{align}
  J_{q,R}^{(1),{\rm bare}}(s)
  =\frac{\alpha_s C_F}{4\pi}
  \Bigg\{&
  \left[\frac4{\epsilon^2}+\frac3\epsilon
  +(7-\pi^2)\right]\delta(s)
  -\left(\frac4\epsilon+3\right)
  \left(\frac1s\right)_{\!*}^{[\mu^2]}
  +4\left(\frac{\ln(s/\mu^2)}s\right)_{\!*}^{[\mu^2]}
  \Bigg\}.
  \label{eq:inclusiveRealBare}
\end{align}
With the virtual contribution in Eq.~\eqref{eq:JvirtualScaleless} vanishing,
Eq.~\eqref{eq:inclusiveRealBare} contains the endpoint poles generated by the
real distribution.  Consequently, the measured bare result through one loop is
\begin{align}
  J_{\rm TEC}^{q,{\rm bare}}(s,\tau)
  &=\left[\delta(s)+J_{q,R}^{(1),{\rm bare}}(s)\right]
  \delta(\tau)
  +\frac{\alpha_s C_F}{2\pi}\mathcal K(s,\tau)
  +\order(\alpha_s^2).
  \label{eq:JbareDecomposition}
\end{align}
Subtracting the ordinary quark-jet counterterm then gives
\begin{align}
  J_{\rm TEC}^{q}(s,\tau,\mu)
  &=J_q(s,\mu)\delta(\tau)
  +\frac{\alpha_s C_F}{2\pi}
  \left[\frac1{s^2}g\!\left(\frac\tau s\right)\right]_{\tau+}
  +\order(\alpha_s^2),
  \label{eq:JoneLoop}
\end{align}
where
\begin{align}
  J_q(s,\mu)
  &=\delta(s)+\frac{\alpha_s C_F}{4\pi}
  \left[(7-\pi^2)\delta(s)
  +4\left(\frac{\ln(s/\mu^2)}s\right)_{\!*}^{[\mu^2]}
  -3\left(\frac1s\right)_{\!*}^{[\mu^2]}
  \right]
  +\order(\alpha_s^2).
  \label{eq:Jinclusive}
\end{align}
Eq.~\eqref{eq:JoneLoop} obeys the reduction relation Eq.~\eqref{eq:inclusiveSum} by construction.

\begin{figure}[t]
  \centering
  \includegraphics[width=0.72\linewidth]{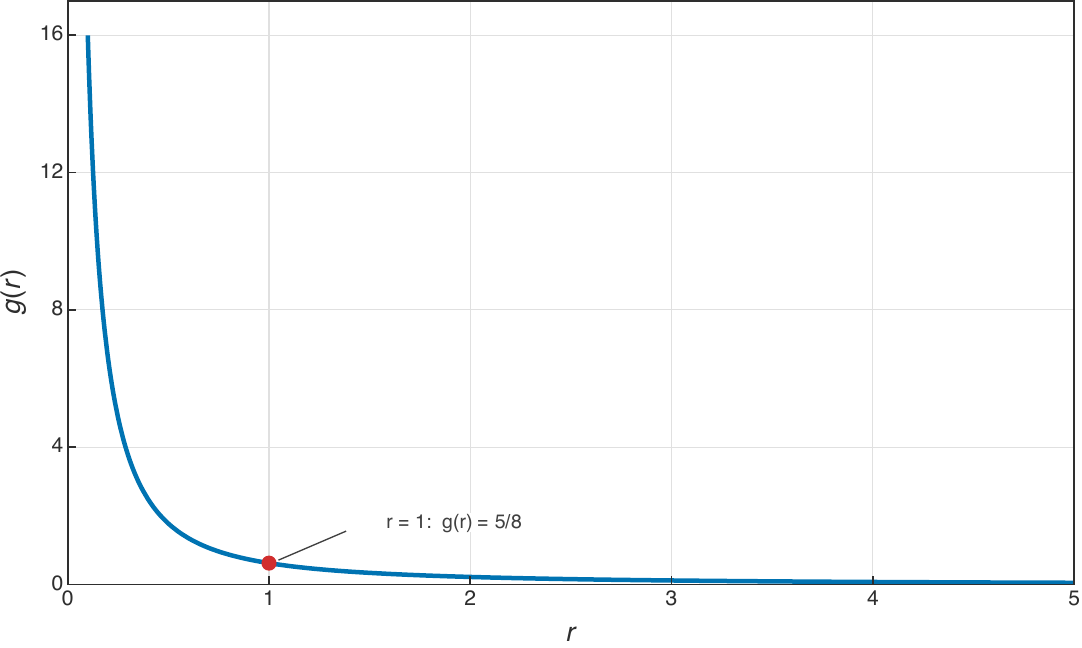}
  \caption{The dimensionless one-loop angular kernel $g(r)$ as a function of
  $r=\tau/s$.  The curve is shown for $0.1\leq r\leq5$
  because $g(r)\sim2/r$ is singular at the origin; at large $r$ it falls as
  $2/r^2$.  The marked point is $g(1)=5/8$.}
  \label{fig:rgr}
\end{figure}

The nontrivial angular dependence is therefore encoded in the dimensionless
function $g(r)$, with $r=\tau/s$.  Its plot is displayed in Fig.~\ref{fig:rgr}.

\section{Renormalization of the jet function}
\label{sec:rg}

At one loop, the ultraviolet poles of the measured jet function are proportional to $\delta(\tau)$ and coincide with those of the inclusive quark jet function. This can be indicated from 
Eq.~\eqref{eq:JbareDecomposition}: the two-dimensional kernel $\mathcal K$ is finite, while every $1/\epsilon^2$ and $1/\epsilon$ pole multiplies $\delta(\tau)$.

In the convention $J_q^{\rm bare}=Z_J^q\otimes_s J_q$, the one-loop $\overline{\rm MS}$ counterterm is
\begin{align}
  Z_J^q(s,\mu)
  &=\delta(s)+\frac{\alpha_s C_F}{4\pi}
  \left\{
  \left(\frac{4}{\epsilon^2}+\frac{3}{\epsilon}\right)\delta(s)
  -\frac{4}{\epsilon}\left(\frac1s\right)_{\!*}^{[\mu^2]}
  \right\}
  +\order(\alpha_s^2).
  \label{eq:ZJexplicit}
\end{align}
The corresponding subtraction for the measured function is
\begin{align}
  J_{\rm TEC}^{q,{\rm bare}}(s,\tau)
  =\int_0^\infty\dd s'\,
  Z_J^q(s-s',\mu)J_{\rm TEC}^{q}(s',\tau,\mu).
  \label{eq:renorm}
\end{align}
To verify Eq.~\eqref{eq:renorm} order by order, insert Eq.~\eqref{eq:JoneLoop} on its right-hand side:
\begin{align}
  Z_J^q\otimes_s J_{\rm TEC}^q
  &=(Z_J^q\otimes_s J_q)\delta(\tau)
  +\frac{\alpha_s C_F}{2\pi}
  (Z_J^{q(0)}\otimes_s\mathcal K)
  +\order(\alpha_s^2)
  \nonumber\\
  &=J_q^{\rm bare}(s)\delta(\tau)
  +\frac{\alpha_s C_F}{2\pi}\mathcal K(s,\tau)
  +\order(\alpha_s^2),
  \label{eq:renormCheck}
\end{align}
where $Z_J^{q(0)}(s)=\delta(s)$.
The convolution of the one-loop part of $Z_J^q$ with $\mathcal K$ starts at $\order(\alpha_s^2)$.
Eq.~\eqref{eq:renormCheck} reproduces Eq.~\eqref{eq:JbareDecomposition} and shows explicitly why the one-loop renormalization kernel is local in $\tau$.
An all-order extension is expected from the energy-weighted inclusiveness of the measurement, but a proof would require an operator-level analysis of the energy-flow insertion and its momentum sum rule.

The one-loop renormalization group equation (RGE) is
\begin{align}
  \mu\frac{\dd}{\dd\mu}J_{\rm TEC}^{q}(s,\tau,\mu)
  =\int_0^\infty\dd s'\,
  \gamma_J^q(s-s',\mu)J_{\rm TEC}^{q}(s',\tau,\mu),
  \label{eq:RGE}
\end{align}
with
\begin{align}
  \gamma_J^q(s,\mu)
  =\frac{\alpha_s C_F}{4\pi}
  \left[-8\left(\frac1s\right)_{\!*}^{[\mu^2]}
  +6\delta(s)\right]
  +\order(\alpha_s^2).
  \label{eq:gammaJ}
\end{align}
After setting $u=\mu^2$, the scale derivatives of the star distributions are
\begin{align}
\begin{aligned}
  \mu\frac{\dd}{\dd\mu}
  \left(\frac1s\right)_{\!*}^{[\mu^2]}
  &=-2\delta(s),
  \\
  \mu\frac{\dd}{\dd\mu}
  \left(\frac{\ln(s/\mu^2)}s\right)_{\!*}^{[\mu^2]}
  &=-2\left(\frac1s\right)_{\!*}^{[\mu^2]}.
\end{aligned}
\label{eq:starScaleDerivatives}
\end{align}
Using the first relation in Eq.~\eqref{eq:starScaleDerivatives} together with
$\mu\,\dd\alpha_s/\dd\mu=-2\epsilon\alpha_s+\order(\alpha_s^2)$ in Eq.~\eqref{eq:ZJexplicit} cancels the remaining $1/\epsilon$ terms and gives Eq.~\eqref{eq:gammaJ} from
$\gamma_J^q=-(Z_J^q)^{-1}\otimes_s\mu\,\dd Z_J^q/\dd\mu$.
This RGE resums renormalization logarithms of $s/\mu^2$ but leaves the
finite angular kernel unchanged at order $\alpha_s$.

\section{Cumulative observables and phenomenological applications}
\label{sec:cumulative}

\subsection{Cumulative observables}

We now turn to cumulative observables derived from the factorization formula
and the measured jet function.  The finite angular kernel can be integrated
analytically.  We define its upper tail by
\begin{align}
  G(R)
  \equiv\int_R^\infty\dd r\,g(r)
  =2\ln\frac{1+R}{R}-\frac{3}{2(1+R)^2},
  \label{eq:gTail}
\end{align}
where $R=\tau_c/s>0$.
For a boundary well below the jet scale, Eq.~\eqref{eq:gTail} becomes
\begin{align}
  G(R)
  =2\ln\frac1R-\frac32+\order(R)
  =2\ln\frac{s}{\tau_c}-\frac32
  +\order\!\left(\frac{\tau_c}{s}\right).
  \label{eq:gTailSmallR}
\end{align}
This logarithm is generated by integrating the small-$r$ behavior
$g(r)\sim2/r$ from $r=R$ to $r\sim1$. 
The perturbative near-axis hierarchy is
$  \Lambda_{\rm QCD}^2\ll\tau_c\ll s$.
If $\tau_c$ approaches $\Lambda_{\rm QCD}^2$, the boundary resolves soft
transverse momentum and the
one-dimensional shape function must be replaced by a
transverse-momentum-dependent soft--collinear factorization.  We leave that
smaller-angle regime for future work.  Setting $f=1$ and
$\tau_{\max}=\tau_c$ in Eq.~\eqref{eq:tauplusFinite} gives
\begin{align}
  &\int_0^{\tau_c}\dd\tau
  \left[\frac1{s^2}g\!\left(\frac\tau s\right)\right]_{\tau+}
  =-\frac1sG\!\left(\frac{\tau_c}{s}\right),
  \label{eq:cumulativeKernel}
\end{align}
where the minus sign follows from the half-line plus prescription.
Eq.~\eqref{eq:cumulativeKernel} therefore isolates the one-loop angular weight above the cut $\tau=\tau_c$.

This observation motivates the fixed-energy near-angle fraction
\begin{align}
  F_{\rm near}(E_\gamma;\tau_c)
  =\frac{\displaystyle\int_0^{\tau_c}\dd\tau\,
  \frac{\dd^2\Gamma_{\BTEC}^{s\gamma}}{\dd E_\gamma\dd\tau}}
  {\displaystyle\int_0^{Q^2}\dd\tau\,
  \frac{\dd^2\Gamma_{\BTEC}^{s\gamma}}{\dd E_\gamma\dd\tau}}.
  \label{eq:Fnear}
\end{align}
At leading power, the denominator can be extended from the physical boundary
$Q^2$ to the auxiliary collinear support,  the inclusive sum rule
in Eq.~\eqref{eq:inclusiveSum} then reduces it to the ordinary photon spectrum,
up to corrections of order $s/Q^2$.  Thus $F_{\rm near}$ is the fraction of
the photon-spectrum bin retained below the angular boundary.

We choose the outer boundary in the natural collinear range,
$\tau_c\sim Q\Lambda_{\rm QCD}$.  The differential relation does not apply pointwise at
$\tau\sim\Lambda_{\rm QCD}^2$, where a transverse-momentum-dependent
factorization would be required; however, no extrapolation into this region is needed. To see this,
note that the exact inclusive relation in Eq.~\eqref{eq:fullQCDInclusiveRelation} gives
\begin{align}
  \int_0^{\tau_c}\dd\tau\,
  \frac{\dd^2\Gamma_{\BTEC}^{s\gamma}}
       {\dd E_\gamma\dd\tau}
  =\frac{\dd\Gamma^{s\gamma}}{\dd E_\gamma}
  -\int_{\tau_c}^{Q^2}\dd\tau\,
  \frac{\dd^2\Gamma_{\BTEC}^{s\gamma}}
       {\dd E_\gamma\dd\tau}.
  \label{eq:cumulativeInclusiveMinusTail}
\end{align}
The first term on the right hand side is the ordinary photon spectrum and includes the complete
leading contribution from the unresolved region, including the Born term and
its nonperturbative transverse broadening.  This contribution is not power
suppressed.  The second term starts at $\tau_c\sim s$ and is dominated by the
natural collinear region, where the measured jet function applies.  The
wide-angle end of the tail contributes at relative order $s/Q^2$.
Consequently, Eq.~\eqref{eq:Fnear} is calculable at leading power without a
pointwise description of $\tau\sim\Lambda_{\rm QCD}^2$.  The remaining
corrections scale as $\Lambda_{\rm QCD}^2/s$ from the invariant-mass shift and
$\Lambda_{\rm QCD}^2/\tau_c$ from soft-recoil migration across the cut, as
discussed in Appendix~\ref{app:transverse}.  For
$\tau_c\sim s\sim Q\Lambda_{\rm QCD}$, all three corrections are of relative
order $\Lambda_{\rm QCD}/Q$.

Unbounded positive moments over the full angular range are not described by
the leading factorization relation because their growing weights promote
wide-angle soft radiation.  In particular, Eq.~\eqref{eq:firstMomentIdentity}
shows that the first moment carries essentially the same information as
$M_X^2$.  We therefore focus on the differential spectrum and bounded
cumulative or annular observables.

To compare with data, we integrate over a photon-energy bin
$\mathcal E_i$ and normalize to the corresponding inclusive spectrum,
\begin{align}
  F_i(\tau_c)
  =\frac{
  \displaystyle\int_{\mathcal E_i}\dd E_\gamma
  \int_0^{\tau_c}\dd\tau\,
  \frac{\dd^2\Gamma_{\BTEC}^{s\gamma}}
       {\dd E_\gamma\dd\tau}}
  {\displaystyle\int_{\mathcal E_i}\dd E_\gamma\,
  \frac{\dd\Gamma^{s\gamma}}{\dd E_\gamma}},
  \qquad \tau_c=\order(Q\Lambda_{\rm QCD}).
  \label{eq:binnedCumulative}
\end{align}
The same fixed $\tau_c$ is applied to every photon-energy bin, so that the
bins are compared at a common physical angular boundary.  The difference
$F_i(\tau_{c,2})-F_i(\tau_{c,1})$ gives the normalized weight in the
corresponding annulus.  At leading power, the hard coefficient and the
overall normalization cancel in Eq.~\eqref{eq:binnedCumulative}; as a result, the remaining
dependence is carried by the shape function and the measured jet kernel.

\subsection{Illustrative numerical benchmark}

To illustrate the resolved angular tail in
Eq.~\eqref{eq:cumulativeInclusiveMinusTail}, we evaluate
Eq.~\eqref{eq:binnedCumulative} in a transparent benchmark rather than a
precision fit.  The arguments of the cumulative kernel are
\begin{align}
  s&=Q(P_+-\widehat\omega),
  \qquad
  R(P_+,\widehat\omega;\tau_c)
  =\frac{\tau_c}{s}
  =\frac{\tau_c}{Q(P_+-\widehat\omega)},  \qquad
   \tau_c=\order(Q\Lambda_{\rm QCD}).
  \label{eq:benchmarkCutKinematics}
\end{align}
We set $Q=P_-=m_B$ in the numerical evaluation.  For the nonperturbative
input, we adopt the two-component model for the leading shape function introduced in
Ref.~\cite{Bosch:2004th},
\begin{align}
  \widehat S_{\rm BLNP}(\widehat\omega,\mu_i)
  ={}&\frac{N}{\Lambda}
  \left(\frac{\widehat\omega}{\Lambda}\right)^{b-1}
  e^{-b\widehat\omega/\Lambda}
-\frac{C_F\alpha_s(\mu_i)}{\pi}
  \frac{\theta(\widehat\omega-\Lambda-\mu_i/\sqrt{e})}
  {\widehat\omega-\Lambda}
  \left[2\ln\frac{\widehat\omega-\Lambda}{\mu_i}+1\right],
  \label{eq:shapeBenchmark}
\end{align}
where
\begin{align}
  N=\left[1-\frac{C_F\alpha_s(\mu_i)}{\pi}
  \left(\frac{\pi^2}{24}-\frac14\right)\right]
  \frac{b^b}{\Gamma(b)}.
  \label{eq:shapeBenchmarkNorm}
\end{align}
We identify
$\widehat S_B^{\rm mod}(\widehat\omega)
=\widehat S_{\rm BLNP}(\widehat\omega,\mu_i)$ and use
$\mu_i=1.5~{\rm GeV}$.
Our central input is model S5 of Table~1 in Ref.~\cite{Bosch:2004th},
$\Lambda=0.685~{\rm GeV}, b=2.93$,
corresponding to $m_b^{\rm SF}=4.65~{\rm GeV}$,
$\bar\Lambda=0.63~{\rm GeV}$, and
$\mu_\pi^2=0.27~{\rm GeV}^2$ in the shape-function scheme.
We estimate the model dependence with the published S1--S9 parameter sets, which span $0.611\leq\Lambda\leq0.759~{\rm GeV}$ and $1.92\leq b\leq4.40$ in correlated pairs.
Although the intrinsic support extends to infinity, the jet constraint limits
the benchmark convolution to
$0\leq\widehat\omega\leq P_+\leq0.9~{\rm GeV}$.
This interval lies below the onset $\Lambda+\mu_i/\sqrt e$ of the radiative
tail for all model choices used below.
The benchmark therefore probes the generalized exponential core of the full BLNP model, and the overall factor $N$ cancels in each normalized cumulative.
For a photon-energy bin $\mathcal E_i=[E_i^{\rm min},E_i^{\rm max}]$,
define its image under $P_+(E_\gamma)=m_B-2E_\gamma$ by $ \mathcal B_i
  =[m_B-2E_i^{\rm max},m_B-2E_i^{\rm min}]$.
The linear change of variables gives $|\dd E_\gamma|=\dd P_+/2$, so this
common Jacobian cancels in each
normalized cumulative.  We therefore evaluate the benchmark in the derived
interval $\mathcal B_i$.  

Let $\mathcal N_i$ and $\mathcal D_i$ denote the
numerator and denominator in the photon-energy bin $\mathcal E_i$, evaluated
with the common fixed cut in Eq.~\eqref{eq:benchmarkCutKinematics}.
Writing $a_J=\alpha_s(\mu_J)C_F/(2\pi)$, their perturbative expansions take the form
\begin{align}
  \mathcal N_i
  &=\mathcal N_i^{(0)}+a_J\mathcal N_i^{(1)}
    +\order(a_J^2),
  &
  \mathcal D_i
  &=\mathcal D_i^{(0)}+a_J\mathcal D_i^{(1)}
    +\order(a_J^2).
  \label{eq:benchmarkPerturbativeExpansion}
\end{align}
The tree-level measurement is proportional to $\delta(\tau)$, and hence any positive cut contains the complete Born contribution.
After canceling the common hard and kinematic factors, one has
\begin{align}
  \mathcal N_i^{(0)}=\mathcal D_i^{(0)}
  =\int_{\mathcal B_i}\dd P_+\,\widehat S_B^{\rm mod}(P_+).
  \label{eq:benchmarkBornBins}
\end{align}
Using Eq.~\eqref{eq:cumulativeKernel}, the one-loop numerator can be written as
the inclusive coefficient minus a positive out-of-cut migration term:
\begin{align}
  \mathcal N_i^{(1)}
  &=\mathcal D_i^{(1)}-\mathcal M_i(\tau_c),
 \nonumber \\
  \mathcal M_i(\tau_c)
  &\equiv
  \int_{\mathcal B_i}\dd P_+
  \int_0^{P_+}\dd\widehat\omega\,
  \frac{\widehat S_B^{\rm mod}(\widehat\omega)}
  {P_+-\widehat\omega}
  G\!\left(\frac{\tau_c}{Q(P_+-\widehat\omega)}\right).
  \label{eq:benchmarkMigrationIntegral}
\end{align}
Consequently, the ratio is expanded before numerical evaluation,
\begin{align}
  \frac{\mathcal N_i}{\mathcal D_i}
  &=\frac{\mathcal D_i^{(0)}
  +a_J[\mathcal D_i^{(1)}-\mathcal M_i(\tau_c)]}
  {\mathcal D_i^{(0)}+a_J\mathcal D_i^{(1)}}
\nonumber
  \\
  &=1-a_J\frac{\mathcal M_i(\tau_c)}{\mathcal D_i^{(0)}}
  +\order(a_J^2).
  \label{eq:benchmarkRatioExpansion}
\end{align}
Substituting the explicit expressions for $\mathcal M_i$ and $\mathcal D_i^{(0)}$ gives
\begin{align}
  F_i^{\rm NLO}(\tau_c)
  &=1-\frac{\alpha_s(\mu_J) C_F}{2\pi}
  \frac{\displaystyle
  \int_{\mathcal B_i}\dd P_+
  \int_0^{P_+}\dd\widehat\omega\,
  \frac{\widehat S_B^{\rm mod}(\widehat\omega)}
  {P_+-\widehat\omega}
  G\!\left(\frac{\tau_c}{Q(P_+-\widehat\omega)}\right)}
  {\displaystyle
  \int_{\mathcal B_i}\dd P_+\,\widehat S_B^{\rm mod}(P_+)}
  +\order(\alpha_s^2).
  \label{eq:numericalCumulative}
\end{align}
Note that the apparent endpoint at $\widehat\omega=P_+$ is regular because
$G(R)\sim2/R$ for $R\to\infty$.

Table~\ref{tab:benchmark} gives the result for $\alpha_s(\mu_J)=0.30$.
The fixed cuts $\tau_c=2$, $3$, and $4~{\rm GeV}^2$ lie in the common
collinear range of the three photon-energy bins.  The benchmark does not
predict the pointwise spectrum at $\tau\sim\Lambda_{\rm QCD}^2$; its complete
integrated weight is supplied by the inclusive term in
Eq.~\eqref{eq:cumulativeInclusiveMinusTail}.
\begin{center}
  \refstepcounter{table}\label{tab:benchmark}
  \begin{minipage}{0.92\linewidth}
  \small
  \textbf{TABLE~\thetable.} Illustrative normalized cumulatives from Eq.~\eqref{eq:numericalCumulative} for the central BLNP model S5 and $\alpha_s(\mu_J)=0.30$.
  The photon-energy bins and their equivalent $P_+$ intervals are in GeV,
  while the fixed values of $\tau_c$ are in ${\rm GeV}^2$;
  $m_B=5.28~{\rm GeV}$ is used only for this bin conversion.
  \par\medskip
  \centering
  \begin{tabular}{cc ccc}
    \toprule
    $\mathcal E_i$ & $\mathcal B_i$
    & $\tau_c=2$ & $\tau_c=3$ & $\tau_c=4$ \\
    \midrule
    $[2.39,2.49]$ & $[0.3,0.5]$ & 0.943 & 0.959 & 0.968 \\
    $[2.29,2.39]$ & $[0.5,0.7]$ & 0.901 & 0.927 & 0.942 \\
    $[2.19,2.29]$ & $[0.7,0.9]$ & 0.837 & 0.878 & 0.901 \\
    \bottomrule
  \end{tabular}
  \end{minipage}
\end{center}
\begin{figure}[b]
	\centering
	\includegraphics[width=0.5\linewidth]{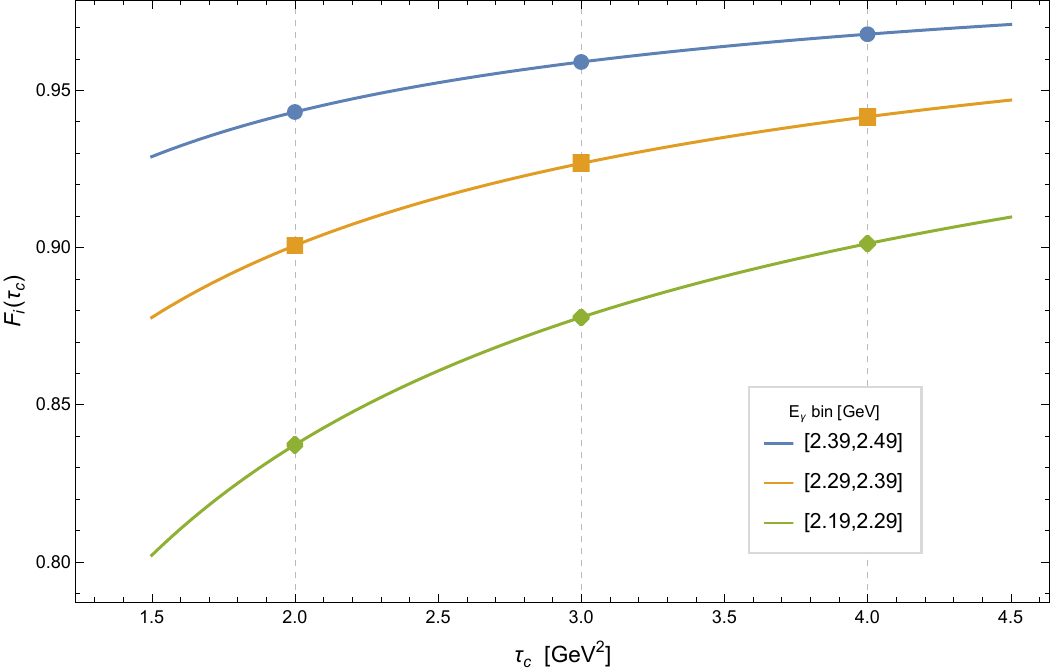}
	\caption{Normalized cumulative $F_i^{\rm NLO}(\tau_c)$ in the three
	photon-energy bins of Table~\ref{tab:benchmark}, evaluated with the central BLNP
	model S5 and $\alpha_s(\mu_J)=0.30$.  The solid curves show the continuous
	dependence on the common fixed cut $\tau_c$.  The filled markers at
	$\tau_c=2$, $3$, and $4~{\rm GeV}^2$ reproduce the corresponding
	entries in Table~\ref{tab:benchmark}, and the dashed vertical lines indicate
	these three tabulated values.}
	\label{fig:fivstauc}
\end{figure}

Fig.~\ref{fig:fivstauc} displays the continuous fixed-cut dependence
underlying Table~\ref{tab:benchmark}.  The cumulative increases as the common
angular boundary is opened and approaches the inclusive normalization.  At
fixed $\tau_c$, the lower photon-energy bins have smaller values of
$F_i(\tau_c)$ and hence larger migration fractions.  This trend directly
reflects the larger recoil mass $M_X^2=QP_+$ and broader jet at lower
$E_\gamma$.

For the representative cut $\tau_c=3~{\rm GeV}^2$, the one-loop kernel places
approximately $4.1\%$, $7.3\%$, and $12.2\%$ of the normalized weight above
the cut across the three bins.
For the combined interval $2.19<E_\gamma<2.49~{\rm GeV}$, equivalently
$0.3<P_+<0.9~{\rm GeV}$, one finds
$F^{\rm NLO}(3~{\rm GeV}^2)=0.926$.
At fixed $\alpha_s(\mu_J)=0.30$, the S1--S9 models give
$0.905<F^{\rm NLO}(3~{\rm GeV}^2)<0.941$ for the combined interval.
Varying in addition $\alpha_s(\mu_J)$ from $0.25$ to $0.35$ broadens this illustrative envelope to
$0.889<F^{\rm NLO}(3~{\rm GeV}^2)<0.951$.
For the three separate bins at this cut, the joint model and coupling
variation corresponds to migration ranges of $2.5$--$6.6\%$,
$4.5$--$11.5\%$, and $7.7$--$18.4\%$, respectively.
This envelope is not a precision uncertainty estimate because it omits
resummation, subleading shape functions, resolved photons, and detector
effects.  Its purpose is only to establish the characteristic size and
kinematic trend of the resolved angular tail: for natural cuts the migration
is a few-to-ten-percent effect and can reach the double-digit level in the
lowest photon-energy bin.

\subsection{Closure test of endpoint factorization}

Beyond the numerical benchmark, the fixed-cut cumulatives provide a closure
test of endpoint factorization.  The photon spectrum and the cumulatives in
Eq.~\eqref{eq:binnedCumulative} depend on the same leading shape function but
probe it with different jet kernels.  We therefore first infer a posterior
$\widehat S_B^{\rm post}$ from the ordinary photon spectrum, including its
correlations with heavy-quark parameters and perturbative scales.  This
posterior is then propagated, without an unrestricted refit to the BTEC data,
to the fixed-cut cumulatives.  The comparison is quantified by
\begin{align}
  \Delta_i(\tau_c)
  \equiv F_i^{\rm data}(\tau_c)
  -F_{i,{\rm LP}}^{\rm pred}(\tau_c;\widehat S_B^{\rm post}).
  \label{eq:closureResidual}
\end{align}
After correlated theoretical and experimental uncertainties are propagated,
leading-power endpoint factorization predicts $\Delta_i(\tau_c)=0$ up to
power corrections.  The test asks whether the shape function calibrated by
the inclusive projection also describes its angularly resolved projections.

Schematically, the first departures from this prediction can be organized as
\begin{align}
  F_i(\tau_c)
  &=F_{i,{\rm LP}}(\tau_c;\widehat S_B^{\rm post})
  +\delta F_{i,{\rm dir}}^{(1/m_b)}(\tau_c)
  +\delta F_{i,{\rm res}\,\gamma}^{(1/m_b)}(\tau_c)
  +\delta F_{i,{\rm kin}}(\tau_c)+\cdots.
  \label{eq:powerCorrectionDecomposition}
\end{align}
The direct term contains subleading SCET currents, Lagrangian insertions, and
subleading shape functions~\cite{Lee:2004ja,Bosch:2004cb}.  The resolved-photon
term arises when the photon couples to light partons and introduces additional
soft and jet functions~\cite{Benzke:2010js,Benzke:2022vki,Bartocci:2024bbf}.  The normalization in
Eq.~\eqref{eq:binnedCumulative} removes the common leading hard coefficient,
but it does not remove the bin-dependent distortions generated by these
subleading contributions.  The kinematic term includes the transverse-momentum
correlation and phase-space corrections omitted from the leading theorem.

The scan in both $E_\gamma$ and the common cut $\tau_c$ supplies the relevant
lever arm.  A power correction can be absorbed into a shifted
$\widehat S_B$ in a spectrum-only fit, but that shift predicts a definite
correlated pattern across angular boundaries.  A correction with a different
angular dependence cannot reproduce all cumulatives simultaneously. At the
level of functional dependence, the test remains insensitive to a correction
that has the same form as a shape-function shift in both the inclusive and
angular projections.

At the level of the present relation, a nonzero residual diagnoses an
incomplete leading-power description but does not identify a unique
subleading mechanism.  Separating the terms in
Eq.~\eqref{eq:powerCorrectionDecomposition} requires their measured
subleading-power factorization formulae and angular kernels.  Higher-order
resummation, detector unfolding, and correlated scale variations must also be
controlled before a residual is interpreted as a hadronic power correction.
Conversely, a residual consistent with zero validates the leading-power
description only in the measured angular projections.

\subsection{Experimental feasibility}

The required event topology is already available in hadronic-tag analyses of
the inclusive photon spectrum at Belle~II
\cite{Belle-II:2018jsg,Belle-II:2022hys}.  Reconstructing the companion $B$
fixes the signal-$B$ rest frame and assigns the remaining particles to the
signal side.  The tagged photon then defines the recoil axis event by event.
The calorimetric BTEC only adds an energy-weighted angular sum over the
remaining signal-side objects; it requires neither an exclusive
reconstruction of $X_s$ nor a jet algorithm.  These features make it a
natural extension of the existing tagged photon-spectrum measurement.

Fixed-cut cumulatives are the most robust first observables.  Their integrated
nature avoids resolving the detailed spectrum in the nonperturbative
small-$\tau$ region, and the energy weight reduces the influence of very soft
objects.  Normalizing to the photon spectrum in the same $E_\gamma$ bin also
allows common tagging, luminosity, and photon-selection uncertainties to
cancel partially.  Detector effects do not disappear; they enter mainly
through migration across the $E_\gamma$ bin edges and the boundary $\tau_c$.
The normalization sum rule in Eq.~\eqref{eq:inclusiveSum} and the first-moment
identity in Eq.~\eqref{eq:firstMomentIdentity} provide independent checks of
the reconstructed energy flow.

A minimal analysis strategy is therefore sufficient.  The
double-differential spectrum is unfolded in $(E_\gamma,\tau)$ using a fixed
particle-level convention, with $Q=P_-=m_B$ and
$P_+(E_\gamma)=m_B-2E_\gamma$.  The same fixed $\tau_c$ is then applied in
every photon-energy bin.  We first constrain $\widehat S_B$ with the ordinary
photon spectrum and use it, without refitting, to predict the fixed-cut
cumulatives.  The full covariance between the two measurements is retained.
If $\widehat S_B$ were allowed to vary freely in the BTEC fit, deviations from
leading-power factorization could instead be absorbed into the fitted shape
function.

The main practical limitations would be the efficiency of hadronic tagging,
neutral-energy reconstruction, detector acceptance, and signal-side particle
assignment.  Their impact on migration across $\tau_c$ must be quantified
with detector simulation and control samples before the few-to-ten-percent
effect in Table~\ref{tab:benchmark} can be interpreted.  Separating charged
and neutral tags is experimentally natural and provides an additional handle
on spectator-dependent resolved-photon effects.  Measurements restricted to
charged tracks or identified hadrons define different observables.  Their
factorization requires track functions or transverse-momentum-dependent
fragmentation functions~\cite{Chang:2013rca,Jaarsma:2023ell,Mi:2025abd}, and
they cannot be treated as the calorimetric BTEC with unobserved particles
simply removed.

\subsection{Semileptonic extension}

Once the radiative construction has been tested, the same measured-jet
framework can be transferred to the endpoint region of
$\Bbar\to X_u\ell\bar\nu$.  At leading power, schematically,
\begin{align}
  \frac{\dd\Gamma_{\BTEC}^{u}}
       {\dd\Phi_L\,\dd P_+\,\dd\tau}
  &=\Gamma_u^{(0)}|V_{ub}|^2H_u(\Phi_L,m_b,\mu)P_-
  \int_0^{P_+}\dd\widehat\omega\,
  J_{\rm TEC}^{q}\!\left(P_-[P_+-\widehat\omega],\tau,\mu\right)
  \widehat S_B(\widehat\omega,\mu)
+\text{power corrections},
  \label{eq:semileptonicFact}
\end{align}
where $\Phi_L$ denotes the leptonic variables.  The same measured light-quark
jet function and leading shape function appear when the recoil direction is
fixed by the nonhadronic system, the observable is azimuthally inclusive, and
transverse-momentum balance remains power suppressed at the natural angular
scale.  Bounded cumulatives may then be formed in bins of $\Phi_L$ and $P_+$,
with the outer cut chosen at $\tau_c\sim P_-\Lambda_{\rm QCD}$.

Using the shape function constrained by the radiative spectrum, the
semileptonic cumulatives test its leading-power universality
\cite{Belle-II:2025pye}.  If this test is successful, the same observables
could provide additional constraints on endpoint dynamics and shape-function
uncertainties in an inclusive $|V_{ub}|$ extraction.  Experimentally, this
requires reconstruction of the nonhadronic recoil direction and control of
the dominant charm background.  The present factorization formula predicts the angular distribution of the
$X_u$ signal, but not that of the dominant $X_c$ background.  The latter can
be measured in charm-enriched control samples, allowing the difference
between the $X_u$ and $X_c$ angular profiles to be used in their separation.
A combined fit to the photon-energy and angular distributions may therefore
improve the inclusive extraction of $|V_{ub}|$.

\section{Summary}
\label{sec:summary}

We have developed a photon-tagged one-point energy correlator for inclusive endpoint $B$ decays.
For the direct-photon contribution to $\Bbar\to X_s\gamma$, the leading spectrum factorizes into the standard hard coefficient, the same universal hatted shape function $\widehat S_B$ as in the photon spectrum, and a new measured quark jet function $J_{\rm TEC}^q$.
The factorization relation applies at the natural collinear angular scale and
breaks down in the parametrically smaller region
$\tau\sim\Lambda_{\rm QCD}^2$, where soft transverse momentum becomes leading
and TMD-like shape information is required.  Bounded cumulatives with an
outer boundary $\tau_c\sim Q\Lambda_{\rm QCD}$ remain leading-power
observables. The angular effect enters only through migration across that
boundary, while the accompanying invariant-mass shift is likewise suppressed
by $\Lambda_{\rm QCD}/Q$.

We calculated $J_{\rm TEC}^q$ at one loop.
Its $\delta(\tau)$ term is fixed by the inclusive quark jet function, while the finite kernel $g(\tau/s)$ contains the new angular information.
The result satisfies the inclusive sum rule and ordinary quark-jet renormalization, and its angular plus prescription requires no additional joint endpoint subtraction.
The analytic cumulative kernel exhibits the boundary logarithm
$\ln(s/\tau_c)$ when $\tau_c\ll s$ and provides the direct perturbative input
for bounded near-angle observables; this logarithm is separate from the
renormalization-group evolution in $\mu$.
The exact event-level first-moment relation reduces to $M_X^2$ at leading power, so the independent information resides in the differential spectrum and bounded cumulatives.

For the central BLNP shape-function model, our illustrative one-loop benchmark finds a $3$--$16\%$ migration outside the common cuts $2\leq\tau_c\leq4~{\rm GeV}^2$ and a clear dependence on the photon-energy bin.
We therefore propose binned cumulative and annular fractions for tagged
$B\to X_s\gamma$ samples.  Rather than treating the BTEC as a second
extraction of the leading shape function, the spectrum-constrained
$\widehat S_B$ is propagated to the angular bins in a closure test of
leading-power endpoint factorization.  We also assess the feasibility of a
calorimetric measurement and give the leading-power extension to semileptonic decay,
where the radiative calibration provides a baseline for testing
shape-function universality before an inclusive $|V_{ub}|$ application.
The next quantitative step is a covariance-aware closure analysis including correlated scale variations, subleading shape functions, resolved-photon effects, experimental response, and power-correction nuisance parameters.
More broadly, the angular information resolves the internal energy flow of
the inclusive recoil jet beyond the ordinary spectrum and may improve the
separation of signals from backgrounds with different angular profiles.

\begin{acknowledgments}
	This work was supported in part by the National Natural Science Foundation
	of China under Contract No.~12475098.
\end{acknowledgments}

\appendix

\section{Transverse-momentum reduction and its accuracy}
\label{app:transverse}

This appendix concernsthe momentum-conservation delta function in
Eq.~\eqref{eq:fourDeltaLightCone}.  The  issue here is whether
the soft transverse momentum must be retained in the transverse component of
overall momentum conservation.  Note that it is not needed to replace
$\calT_X$ by $\calT_{X_n}$ because that measurement reduction follows directly from
Eq.~\eqref{eq:directSoftMeasurementBound}.

The cleanest argument is obtained by integrating the transverse delta
function exactly.  Let
$\mathcal J_{\rm TEC}^q(s,\tau,\boldsymbol p_\perp)$ denote the same measured
collinear cut as in Eq.~\eqref{eq:Jdefinition}, but before its total
transverse momentum is set to zero.  The standard measured jet function is
therefore
\begin{align}
  J_{\rm TEC}^q(s,\tau)
  =\mathcal J_{\rm TEC}^q(s,\tau,\boldsymbol 0_\perp).
  \label{eq:unprojectedJetRelation}
\end{align}
Similarly, let
$\widehat S_B(\widehat\omega,\boldsymbol k_\perp)$ denote the soft state sum
differential in its total transverse momentum.  Its transverse integral is
the ordinary shape function,
\begin{align}
  \widehat S_B(\widehat\omega)
  =\int\dd^2\boldsymbol k_\perp\,
  \widehat S_B(\widehat\omega,\boldsymbol k_\perp).
  \label{eq:TMDShapeIntegral}
\end{align}
After the transverse delta function is integrated, the corresponding part of
the exact state sum has the schematic function-level form
\begin{align}
  W_\perp(s,\tau,\widehat\omega)
  \sim\int\dd^2\boldsymbol k_\perp\,
  \widehat S_B(\widehat\omega,\boldsymbol k_\perp)\,
  \mathcal J_{\rm TEC}^q
  (s,\tau,-\boldsymbol k_\perp).
  \label{eq:TMDLikeConvolution}
\end{align}
For compactness, Eq.~\eqref{eq:TMDLikeConvolution} displays $s$ as an
independent argument.  The accompanying
$\order(\boldsymbol k_\perp^2)$ shift of its physical value is shown
explicitly in Eq.~\eqref{eq:sargument} and is included in the estimate below.
Before the estimation, note that two things can be learned from
Eq.~\eqref{eq:TMDLikeConvolution}: (1) The
exact constraint gives
$p_{X_n\perp}=-k_{X_s\perp}\sim\Lambda_{\rm QCD}$.  The larger scale
$Q\lambda$ belongs to transverse motion inside the collinear state and sets
the resolution of the collinear function; it is not the value of the total
momentum on the support of the delta function.  (2) Recovering the
standard product of Eqs.~\eqref{eq:unprojectedJetRelation} and
\eqref{eq:TMDShapeIntegral} requires the collinear function to be insensitive
to a total displacement of order $k_{X_s\perp}$.

The $B$ meson is spinless, and the BTEC does not retain an azimuth about the
photon axis.  Transverse rotational invariance therefore permits
$\mathcal J_{\rm TEC}^q$ to depend on its total transverse argument only
through $\boldsymbol p_\perp^2$.  
For a smooth spectrum or a
finite angular bin, its ordinary function expansion is therefore
\begin{align}
  \mathcal J_{\rm TEC}^q(s,\tau,-\boldsymbol k_\perp)
  =\mathcal J_{\rm TEC}^q(s,\tau,\boldsymbol 0_\perp)
  +\boldsymbol k_\perp^2
  \left.
  \frac{\partial\mathcal J_{\rm TEC}^q(s,\tau,\boldsymbol p_\perp)}
       {\partial\boldsymbol p_\perp^2}
  \right|_{\boldsymbol p_\perp=\boldsymbol 0_\perp}
  +\cdots .
  \label{eq:transverseFunctionExpansion}
\end{align}
There is no term linear in $\boldsymbol k_\perp$.  This is the step that
removes a possible $\order(\lambda)$ correction from the azimuthally
inclusive rate.  A bin at $\tau$ resolves constituent transverse momentum of
order $\sqrt{\tau}$, so the derivative in
Eq.~\eqref{eq:transverseFunctionExpansion} is controlled by the squared scale
$\tau$.  In addition, retaining the exact total transverse momentum shifts
the collinear invariant mass by $\order(\boldsymbol k_\perp^2)$, whose
variation scale is $s$.  With
$\boldsymbol k_\perp^2\sim\Lambda_{\rm QCD}^2$, the two relative corrections
are therefore
\begin{align}
  \frac{\Delta W_\perp}{W_\perp}
  \sim \mathcal{O}\left(\frac{\Lambda_{\rm QCD}^2}{\tau}\right)
  + \mathcal{O}\left(\frac{\Lambda_{\rm QCD}^2}{s}\right).
  \label{eq:appendixTransverseCorrection}
\end{align}
At the natural collinear scale
$\tau\sim s\sim Q\Lambda_{\rm QCD}$, both terms are
$\Lambda_{\rm QCD}/Q$.  This proves the accuracy quoted in the main text
without treating the bare delta function as an ordinary function.

The same reasoning applies directly to a bounded cumulative.  Define the
unprojected cumulative collinear factor by
\begin{align}
  \mathcal J_{<}^q(s,\tau_c,\boldsymbol p_\perp)
  =\int_0^{\tau_c}\dd\tau\,
  \mathcal J_{\rm TEC}^q(s,\tau,\boldsymbol p_\perp).
  \label{eq:unprojectedCumulative}
\end{align}
A transverse displacement redistributes angular weight inside the interval
and changes its angular integral only by moving weight across the outer
boundary.  This part of the expansion is therefore controlled by $\tau_c$,
not by the unresolved point $\tau=0$.  Together with the independent
$\Lambda_{\rm QCD}^2/s$ invariant-mass correction, it is again of relative
order $\Lambda_{\rm QCD}/Q$ for
$\tau_c\sim s\sim Q\Lambda_{\rm QCD}$.  This explains why the cumulatives
in Sec.~\ref{sec:cumulative} remain valid even though their integration range
contains the parametrically smaller-angle region.  In the simplest physical
picture, soft transverse momentum broadens the collinear
$\delta(\tau)$ contribution over a region of width
$\tau\sim\Lambda_{\rm QCD}^2$, therefore, a natural cumulative contains this entire
region and is insensitive to its internal shape.

When $\tau\sim\Lambda_{\rm QCD}^2$, the angular expansion parameter in
Eq.~\eqref{eq:appendixTransverseCorrection} is order one.  The full
convolution in Eq.~\eqref{eq:TMDLikeConvolution} must then be kept, and
$\widehat S_B(\widehat\omega,\boldsymbol k_\perp)$ is a TMD-like shape function
rather than the ordinary one-dimensional
$\widehat S_B(\widehat\omega)$.  Additional small-angle modes may be
required in a complete factorization theorem.  Observables that measure the
transverse recoil vector itself, or processes with a polarized initial state
that supplies additional transverse structure, require a separate power
analysis.

\section{Jet-function normalization and cut phase space}
\label{app:norm}

The cut-state sum in Eq.~\eqref{eq:Jdefinition} includes the Lorentz-invariant phase-space integrations and discrete sums,
\begin{align}
  \sum_X\to\sum_m\frac1{S_m}
  \prod_{i=1}^m\left[
  \sum_{\sigma_i,c_i}
  \int\frac{\dd^d p_i}{(2\pi)^{d-1}}
  \theta(p_i^0)\delta(p_i^2)
  \right].
\end{align}
The projectors
\begin{align}
  \delta(Q-\bar n\cdot p_X)
  \delta^{(d-2)}(p_{X\perp})
  \delta(s-p_X^2)
\end{align}
represent $d$ constraints on the total collinear momentum.
They are related to the usual total-momentum projector by
\begin{align}
  (2\pi)^d\delta^{(d)}(P-p_X)
  &=4\pi Q(2\pi)^{d-1}
  \delta(Q-\bar n\cdot p_X)
  \nonumber\\
  &\quad\times
  \delta^{(d-2)}(p_{X\perp})
  \delta(s-p_X^2),
\end{align}
for $P^-=Q$, $P_\perp=0$, and $P^+=s/Q$.

For a one-quark state,
\begin{align}
  \sum_\sigma u_n(p,\sigma)\bar u_n(p,\sigma)
  =\frac{\bar n\cdot p}{2}\slashed n,
\end{align}
and the one-particle phase space with the fixed labels gives $1/(2Q)$.
The spin trace gives $2Q$ and the normalized color trace gives unity.
Thus $\mathcal N_J=1$ yields Eq.~\eqref{eq:Jtree}.

\end{document}